\documentclass{aa}

\usepackage{rotating}
\usepackage{natbib}
\usepackage{mathtools}
\usepackage{amsmath}
\usepackage[normalem]{ulem}
\usepackage[percent]{overpic}

\usepackage[percent]{overpic}

\usepackage[dvipsnames]{xcolor}

\usepackage{breqn}

\DeclarePairedDelimiter\abs{\lvert}{\rvert}%
\makeatletter
\let\oldabs\abs
\def\abs{\@ifstar{\oldabs}{\oldabs*}}

\begin{document}

\title{Collisional ionisation, recombination and ionisation potential in two-fluid slow-mode shocks: analytical and numerical results.}
\titlerunning{PIP IRIP shocks}

\author{B. Snow\inst{1}, A. Hillier\inst{1}}
\institute{University of Exeter, Exeter, EX4 4QF, UK \email{b.snow@exeter.ac.uk} \label{inst1}}


\abstract 
{Shocks are a universal feature of the lower solar atmosphere which consists of both ionised and neutral species. Including partial ionisation leads to a finite-width existing for shocks, where the ionised and neutral species decouple and recouple. As such, drift velocities exist within the shock that lead to frictional heating between the two species, in addition to the adiabatic temperature changes across the shock.
The local temperature enhancements within the shock alter the recombination and ionisation rates and hence change the composition of the plasma.}
{We study the role of collisional ionisation and recombination in slow-mode partially-ionised shocks. In particular we incorporate the ionisation potential energy loss and analyse the consequences of having a non-conservative energy equation.}
{A semi-analytical approach is used to determine the possible equilibrium shock jumps for a two-fluid model with ionisation, recombination, ionisation potential and arbitrary heating. Two-fluid numerical simulations are performed using the (P\underline{I}P) code. Results are compared to the MHD model and semi-analytic solution.}
{Accounting for ionisation, recombination and ionisation potential significantly alters the behaviour of shocks in both substructure and post-shock regions. In particular, for a given temperature, equilibrium can only exist for specific densities due to the radiative losses needing to be balanced by the heating function. A consequence of the ionisation potential is that a compressional shock will lead to a reduction of temperature in the post-shock region, rather than the increase seen for MHD. The numerical simulations pair well with the derived analytic model for shock velocities.}
{}

\keywords{magnetohydrodynamics (MHD), Sun:chromosphere}

\maketitle

\section{Introduction}


A fundamental feature of the solar atmosphere (and astrophysical systems in general) is shock waves. Shocks can form in a number of ways, for example the natural steepening of upwardly propagating waves (e.g., umbral flashes), or from reconnection events. These events all occur in a magnetised plasma and as such the solar atmosphere can support several different types of shock. For magnetohydrodynamic (MHD) systems shocks can be broadly defined as slow, fast or intermediate depending on the velocity either side of the shock \citep[e.g.,][]{Delmont2011}. Here we focus on the slow-mode shock, which is a fundamental feature of magnetic reconnection \citep{Petschek1964}.

The MHD model works well to study shocks in fully-ionised mediums. However, the lower solar atmosphere consists of both ionised and neutral particles and hence two-fluid effects become important in this region \citep[see the review of ][]{Khomenko2017}. Events such as umbral flashes or Ellerman bombs occur in the lower solar atmosphere and are intrinsically linked to shocks. 

For partially-ionised two-fluid (ion+electron, neutral) systems shocks become more complicated. There is a larger set of characteristic wave speeds including the neutral sound speed, the slow, Alfv\'en, fast plasma speeds, and the slow, Alfv\'en, fast bulk (plasma+neutral) speeds \citep{Snow2020}. An effective set of wave speeds also exists that involves using the collisionallity of the system to determine the slow, Alfv\'en and fast speeds \citep{Soler2013}. In a two-fluid shock, the species decouple and recouple leading to a finite width and large drift velocities in the shock \citep{Hillier2016}. Within this finite shock width, additional shock transitions can form \citep{Snow2019}.  


The rapid changes in local properties in a shock alter the ionisation and recombination rates of the system \citep[e.g.,][]{Carlsson2002}.
Due to the temperature changes in and across a shock, the ionisation and recombination rates change, that has the effect of changing the composition of the medium, increasing or decreasing the neutral fraction. The most common way to model the rates of collisional ionisation and recombination is to use either empirical fits to lab data \citep{Smirnov2003,Voronov1997} or a multi-level atom models \citep{Jefferies1968}. The empirical formulas assume a statistical equilibrium where the hydrogen species ionises as an ensemble single fluid and does not account for different levels of excitation within the species itself. A more complete multi-level model includes non-local thermal equilibrium where the ionisation/recombination rates of individual hydrogen levels are modelled separately \citep[e.g.,][]{Carlsson2002}.



During collisional ionisation, kinetic energy of a free electron is used to release an electron that is trapped in the electric field of a nucleus. The macroscopic effect of the work done to ionised an atom is the reduction in internal energy (and with it the temperature) of the plasma fluid. However, as photons can be released as part of the process of collisional recombination/de-excitation the recombination-ionisation cycle of an atom does not necessarily conserve the energy of the fluid, potentially resulting in cooling.


In this paper we focus on the solar atmosphere however, partially-ionised shocks are also a common feature of the interstellar medium. When radiative losses are considered, the energy loss within the finite-width of the shock acts to limit the maximum temperature increase \citep{Draine1993}. Radiative losses then cool the post-shock region as further energy is lost, despite an increase in both density and pressure across the shock. 



In this paper we study the role of collisional ionisation and recombination in a reconnection-driven slow mode shock for a two-fluid partially-ionised hydrogen plasma.
Our previous studies neglected ionisation and recombination and focused on the fundamental two-fluid effects \citep{Hillier2016,Snow2019}. Here we expand the model further towards reality, accounting for collisional ionisation, recombination and ionisation potential. We present a semi-analytical method for determining the possible equilibrium shock solutions for a radiating fluid. We then perform and analyse 1D numerical simulations.
The resultant shocks have different behaviour to the MHD solutions due to the non-conservative energy equation. A key feature is that the ionisation potential acts as an energy loss term and cools the system, despite an increase in both pressure and density across the shock interface.

\section{Methodology}

In this paper, we study ionisation, recombination, ionsiation potential and arbitrary heating in the framework of two-fluid (neutral, and ion+electron) model for hydrogen using a semi-analytical approach and numerical simulations. Specifically, in this study, we use the following set of equations:


\begin{gather}
\frac{\partial \rho _{\text{n}}}{\partial t} + \nabla \cdot (\rho _{\text{n}} \textbf{v}_{\text{n}})= \Gamma _{rec} \rho _{\rm p} - \Gamma _{ion} \rho _{\rm n}, \label{eqn:neutral1} \\
\frac{\partial}{\partial t}(\rho _{\text{n}} \textbf{v}_{\text{n}}) + \nabla \cdot (\rho _{\text{n}} \textbf{v}_{\text{n}} \textbf{v}_{\text{n}} + P_{\text{n}} \textbf{I}) \nonumber \\  \hspace{0.5cm} = -\alpha _c \rho_{\text{n}} \rho_{\text{p}} (\textbf{v}_{\text{n}}-\textbf{v}_{\text{p}}) + \Gamma _{rec} \rho _{\rm p} \textbf{v}_{\rm p} - \Gamma _{ion} \rho_{\rm n} \textbf{v}_{\rm n}, \\
\frac{\partial e_{\text{n}}}{\partial t} + \nabla \cdot \left[\textbf{v}_{\text{n}} (e_{\text{n}} +P_{\text{n}}) \right] \nonumber \\  \hspace{0.5cm}= -\alpha _c \rho _{\text{n}} \rho _{\text{p}} \left[ \frac{1}{2} (\textbf{v}_{\text{n}} ^2 - \textbf{v}_{\text{p}} ^2)+ \frac{3}{2} \left(\frac{P_{\rm n}}{\rho_{\rm n}}-\frac{1}{2}\frac{P_{\rm p}}{\rho_{\rm p}}\right) \right] \nonumber \\ \hspace{0.5cm}+ \frac{1}{2} \left( \Gamma _{rec} \rho _{\rm p} \textbf{v}_{\rm p} ^2 - \Gamma _{ion} \rho _{\rm n} \textbf{v}_{\rm n} ^2 \right) \nonumber \\ \hspace{0.5cm} +\frac{1}{ (\gamma-1)} \left( \frac{1}{2} \Gamma _{rec} P_{\rm p} -\Gamma _{ion} P_{\rm n} \right),  \\
e_{\text{n}} = \frac{P_{\text{n}}}{\gamma -1} + \frac{1}{2} \rho _{\text{n}} v_{\text{n}} ^2, \label{eqn:neutral2} \\
\frac{\partial \rho _{\text{p}}}{\partial t} + \nabla \cdot (\rho_{\text{p}} \textbf{v}_{\text{p}}) = - \Gamma _{rec} \rho _{\rm p} + \Gamma _{ion} \rho _{\rm n} \label{eqn:plasma1}\\
\frac{\partial}{\partial t} (\rho_{\text{p}} \textbf{v}_{\text{p}})+ \nabla \cdot \left( \rho_{\text{p}} \textbf{v}_{\text{p}} \textbf{v}_{\text{p}} + P_{\text{p}} \textbf{I} - \textbf{B B} + \frac{\textbf{B}^2}{2} \textbf{I} \right) \nonumber \\  \hspace{0.5cm}= \alpha _c \rho_{\text{n}} \rho_{\text{p}}(\textbf{v}_{\text{n}} - \textbf{v}_{\text{p}}) - \Gamma _{rec} \rho _{\rm p} \textbf{v}_{\rm p} + \Gamma _{ion} \rho_{\rm n} \textbf{v}_{\rm n},
\end{gather}
\begin{gather}
\frac{\partial}{\partial t} \left( e_{\text{p}} + \frac{\textbf{B}^2}{2} \right) + \nabla \cdot \left[ \textbf{v}_{\text{p}} ( e_{\text{p}} + P_{\text{p}}) -  (\textbf{v}_{\rm p} \times \textbf{B}) \times \textbf{B} \right] \nonumber \\  \hspace{0.5cm} =  \alpha _c \rho _{\text{n}} \rho _{\text{p}} \left[ \frac{1}{2} (\textbf{v}_{\text{n}} ^2 - \textbf{v}_{\text{p}} ^2)+ \frac{3}{2} \left(\frac{P_{\rm n}}{\rho_{\rm n}}-\frac{1}{2}\frac{P_{\rm p}}{\rho_{\rm p}}\right) \right] \nonumber \\ \hspace{0.5cm}- \frac{1}{2} \left( \Gamma _{rec} \rho _{\rm p} \textbf{v}_{\rm p} ^2 - \Gamma _{ion} \rho _{\rm n} \textbf{v}_{\rm n} ^2 \right) - \phi_I + A_{heat} \nonumber \\  \hspace{0.5cm} -\frac{1}{ (\gamma-1)} \left( \frac{1}{2} \Gamma _{rec} P_{\rm p} -\Gamma _{ion} P_{\rm n} \right), \label{eqn:ep} \\
\frac{\partial \textbf{B}}{\partial t} - \nabla \times (\textbf{v}_{\text{p}} \times \textbf{B}) = 0, \\
e_{\text{p}} = \frac{P_{\text{p}}}{\gamma -1} + \frac{1}{2} \rho _{\text{p}} v_{\text{p}} ^2, \\
\nabla \cdot \textbf{B} = 0,\label{eqn:plasma2}
\end{gather}
for a charge neutral plasma (subscript $\mbox{p}$) and neutral (subscript $\mbox{n}$) species. The fluid properties are given by density $\rho$, pressure $P$, velocity $\textbf{v}$, magnetic field $\textbf{B}$ and energy $e$. Both species follow ideal gas laws for temperature $T$, namely $T_{\rm n} = \gamma P_{\rm n}/\rho_{\rm n}$ and $T_{\rm p} = \frac{1}{2} \gamma P_{\rm p}/\rho_{\rm p}$.

The species are thermally coupled through the collisional coefficient $\alpha_c$ which is calculated as:
\begin{gather}
    \alpha _c = \alpha _0 \sqrt{\frac{T_{\rm p}+T_{\rm n}}{2}} \sqrt{\frac{1}{T_{init}}}.
\end{gather}
The factor of $\sqrt{\frac{1}{T_{init}}}$ is to normalise the collisional coefficient using the initial temperature $T_{init}$ such that $\alpha_c (t=0) = \alpha _0$.

\subsection{Collisional Ionisation and Recombination}

The collisional (three-body) ionisation ($\Gamma _{ion}$) and recombination ($\Gamma _{rec}$) rates for a hydrogen atom are given by the empirical forms from \cite{Voronov1997} and \cite{Smirnov2003}. In normalised form, these equations are:

\begin{gather}
    \Gamma_{rec} = \frac{\rho_{\rm p}}{\sqrt{T_{\rm p}}} \frac{\sqrt{T_f}}{\xi _{{\rm p}0}} \tau _{IR} = F(T) \rho_{\rm p}, \\
    \Gamma_{ion} = \rho_{\rm p} \frac{\mbox{e} ^{-\chi} \chi ^{0.39} }{0.232 + \chi} \frac{\hat{R}}{\xi _{{\rm p}0}} \tau _{IR} = G(T) \rho_{\rm p}, \\
    \chi = 13.6 \frac{T_f}{T_{e0} T_{\rm p}}, \\
    \hat{R} = \frac{2.91 \times 10 ^{-14}}{2.6 \times 10^{-19}} \sqrt{T_{e0}},
\end{gather}
where $T_f$ is a normalisation factor to ensure the simulation temperature is the reference temperature $T_0$ in Kelvin. $T_{e0}$ is the reference temperature converted to electron volts. A free parameter exists ($\tau _{IR}$) that governs the rate of ionisation and recombination relative to the collisional time. A detailed derivation is included in Appendix \ref{app:rates}


In Equation (\ref{eqn:ep}) for the plasma energy, the recombination term has a factor of $1/2$ which ensures that the ionisation and recombination terms are in equilibrium at $T_{\rm n}=T_{\rm p}$, consistent with \cite{Leake2012,Popescu2019}. Note that the equations used in \cite{Singh2019} are missing this factor, implying that their ionisation/recombination equilibrium exists at $T_{\rm n}=2 T_{\rm p}$, whereas their collisional equilibrium occurs at $T_{\rm n}=T_{\rm p}$, leading to non-trivial equilibrium conditions.

The energy equation for the plasma species includes an additional term to approximate the ionisation potential and the energy removed from the system due to ionisation, denoted in Equation \ref{eqn:ep} as $\phi _I$. We make several assumptions for the ionisation potential:
\begin{itemize}
    \item On recombination only the proton thermal energy is transferred to the neutral fluid. The thermal energy previously held by the recombined electron remains in the electron fluid (and with that the plasma as a whole) as it is transferred to the free electron involved in the three-body recombination.
    \item The electron is assumed to recombine to an arbitrary level of the atom. 
    \item We assume all the other energy involved in the recombination process (coming from the work done on the recombining electron by the proton electric field) is released as a photon.
    \item The medium is assumed to be optically thin.
    \item The electron in any neutral cascades down from higher levels to the ground state by releasing photons.
    \item To ionise the neutral from the ground state through collisions with another electron, work has to be done by the free electron to release the bound electron, resulting in an energy loss. As we assume this happens to electrons in the ground state, the work done is 13.6eV.
\end{itemize}
All these assumptions allow us to avoid the modelling of the internal structure of the atom and have a detailed understanding of the energy balance between thermal and photon energy involved in the recombination process.

In dimensional form, the energy removed from the plasma due to the ionisation potential is calculated as:
\begin{gather}
    \phi _{I,dim} = 13.6 \Gamma _{ion,dim} n _{\rm n}, 
\end{gather}
where $n_{\rm n}$ is the neutral number density, 13.6 is the ionisation energy of hydrogen (in electron volts), and $\Gamma _{ion,dim}$ is the dimensional ionisation rate.

The non-dimensional form used in our simulations is therefore calculated as
\begin{gather}
    \phi _I = \Gamma _{ion} \rho _{\rm n} \hat{\phi},
\end{gather}
where $\hat{\phi}$ is a constant factor such that this ionisation potential is consistent with the rest of the normalisation. If the system is normalised to $c_s=1,\rho_{tot}=1$ then $\hat{\phi} =\frac{13.6}{K_B \gamma T_0}$. If the system is normalised to $V_A=1$, then $\hat{\phi} = \frac{13.6 \beta}{2 K_B T_0}$. 

An arbitrary heating term $A_{heat}$ is also included to obtain initial equilibrium. The form of this is equal to the ionisation potential energy loss using the equilibrium quantities, i.e.,
\begin{gather}
    A_{heat}=\Gamma _{ion} (t=0) \rho_{\rm n} (t=0) \hat{\phi}.
\end{gather}
This heating term is constant throughout the simulation.


\subsection{Ionisation equilibrium conditions}

When ionisation and recombination are studied (with ionisation potential neglected), the ionisation equilibrium can be easily determined from the continuity equation as:
\begin{gather}
    \Gamma _{ion} \rho _{\rm n} = \Gamma _{rec} \rho_{\rm p}, \\
    \Gamma _{ion}/\Gamma _{rec} = \rho _{\rm p}/ \rho _{\rm n}.
\end{gather}
Therefore, the only constraint for ionisation equilibrium is that the ratio of ionisation/recombination rates equals the ratio of plasma/neutral densities. As such, ionisation equilibrium relies on the ratio of densities, and not the specific density values.

Including the ionisation potential has an interesting effect on the equilibrium states than can be achieved by the system. For an equilibrium, we require that the ionisation potential and heating terms balance:
\begin{gather}
    \hat{\phi} \Gamma _{ion} \rho_{\rm n} = \hat{\phi} \Gamma _{ion} (t=0) \rho _{\rm n} (t=0).
\end{gather}
Assuming a constant ionisation energy $\hat{\phi}$, combined with the equilibrium continuity equation 
we find
\begin{gather}
    \Gamma _{ion} \rho_{\rm n} = \Gamma _{rec} \rho_{\rm p} = \Gamma _{ion} (t=0) \rho _{\rm n} (t=0) =\mbox{const.} \label{eqn:gmconst}
\end{gather}

Now, the ionisation and recombination rates can be expressed as a function of temperature and a plasma density:
\begin{gather}
    \Gamma _{ion} = G(T) \rho _{\rm p}, \\
    \Gamma _{rec} = F(T) \rho _{\rm p}.
\end{gather}
Hence Equation \ref{eqn:gmconst} becomes:
\begin{gather}
    G(T)\rho_{\rm p} \rho_{\rm n} = F(T) \rho_{\rm p} ^2 = \mbox{const.} \label{eqn:ircomp}
\end{gather}
This is an interesting result since it states that for a given temperature, an ionisation equilibrium only exists at specific density values (rather than density ratios when the ionisation potential term is neglected).

\begin{figure}
    \centering
    \includegraphics[width=0.95\linewidth,clip=true,trim=0.9cm 8.8cm 0.9cm 8.8cm]{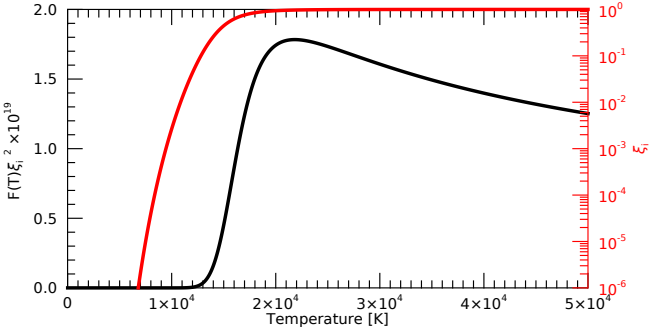}
    \caption{$F(T) \xi_i^2$ as a function of temperature in dimensional form. Red line shows the ionisation fraction $\xi_i$ for the associated temperature.}
    \label{fig:eqltest}
\end{figure}

In terms of bulk density $\rho$, Equation \ref{eqn:ircomp} becomes
\begin{gather}
    F(T) \xi_i ^2 \rho ^2 = \mbox{const.} \label{eqn:irbulk}
\end{gather}
where the ionisation fraction $\xi_i$ can be expressed as a function of temperature only using the conservation of mass equation:
\begin{gather}
    \xi_i=\frac{1}{\frac{F(T)}{G(T)} +1}.
\end{gather}
The behaviour of $F(T) \xi_i ^2$ and $\xi_i$ with respect to temperature are shown in Figure \ref{fig:eqltest}. The shape of $F(T) \xi_i ^2$ implies that there are at most two potential temperatures that satisfy Equation \ref{eqn:ircomp}. In the range $5000<T<20000$ the function in monotonically increasing meaning that a compression of the system requires a decrease in temperature. As such, for a compressible shock one may expect the downstream temperature to be less than the upstream temperature, in contrast to the MHD result where temperature increases across a shock. Above this temperature range, $F(T) \xi_i ^2$ is monotonically decreasing, meaning that a compressible shock has in increase in temperature as is expected for MHD. The change in gradient of the function occurs when the ionisation fraction approaches 1. As such, this equation is consistent with standard MHD theory for a fully ionised plasma, however introduces new behaviour for the temperatures less than $T \approx 20000$ K when the plasma is partially ionised.

In this paper, we study the partially-ionised range for $T<20000$ K. Above this limit, we encounter limitations to our model. A more realistic ionisation energy model would be required to account for the increased likelihood of exciting a higher level when the temperature is higher. Limitations also arise from the numerical model when the neutral density and pressures become too low and the treatment of the neutral species as a fluid breaks down.

\section{Analytic solutions to shock jump equations}

\subsection{MHD}



For the MHD equations, the shock jump conditions can be derived by considering the equations upstream ($^u$) and downstream ($^d$) of the shock in the deHoffmann-Teller shock frame (zero electric field either side of the shock). The conservative nature of the MHD equations mean that for a steady-state shock in the deHoffmann-Teller frame, mass, momentum and energy can be equated sufficiently upstream and downstream. The MHD equations have no source terms and hence can be integrated across the shock to become:  
\begin{gather}
    \left[\rho v_x  \right]^u _d = 0,  \\
    \left[\rho v_x^2 +P +\frac{B_y^2}{2} \right]^u _d = 0, \\
    \left[\rho v_x v_y -B_x B_y \right]^u _d = 0, \\
    \left[ v_{x} \left( \frac{\gamma}{\gamma -1} P + \frac{1}{2} \rho v^2 \right) \right]^u _d =0, \\
    \left[B_x \right]^u _d = 0, \\
    \left[v_x B_y -v_y B_x   \right]^u _d = 0, 
\end{gather}
where this notation means that
\begin{gather}
    \left[ Q \right]^u _d \equiv Q^u - Q^d,
\end{gather}
for any quantity $Q$.

Rearranging these equations, it is possible to derive a single equation for possible shock transitions for a given upstream plasma-$\beta$ and upstream angle of magnetic field $\theta$, that relates the upstream and downstream Alfv\'en Mach numbers \citep{Hau1989}: 
\begin{gather}
    A_x ^{\text{u}2} = \left[ A_x ^{\text{d}2} \left( \frac{\gamma-1}{\gamma} \left( \frac{\gamma+1}{\gamma -1} -\tan ^2 \theta \right) \left(A_x ^{\text{d}2} -1 \right) ^2 \right. \right. \nonumber\\ 
    + \left. \left. \tan ^2 \theta \left( \frac{\gamma-1}{\gamma} A_x ^{\text{d}2} -1 \right) \left(A_x ^{\text{d}2} -2 \right) \right) - \frac{\beta}{ \cos ^2 \theta } \left( A_x ^{\text{d}2} -1 \right) ^2 \right]  \nonumber\\
    / \left[ \frac{\gamma -1}{\gamma} \frac{\left( A_x ^{\text{d}2}-1 \right) ^2}{ \cos ^2 \theta } - A_ x ^{\text{d}2} \tan ^2 \theta \left( \frac{\gamma -1}{\gamma} A_x ^{\text{d}2} -1 \right) \right].
\end{gather}
This is shown in Figure \ref{fig:tjumpex} for a choice of $\beta =0.1, \theta=\pi/4$ for the upstream plasma. A trivial solution exists where the upstream and downstream Alfv\'en Mach numbers are equal. The non-trivial solution intersects the trivial solution at three points where the upstream velocity is equal to the sound speed, the Alfv\'en speed, and the fast speed. Solutions below the trivial solution (where the where the velocity is higher downstream than upstream) are nonphysical because these require that the entropy decreases across the shock. Additional details regarding this MHD solution can be found in \citep{Hau1989,Snow2019}.

In the MHD model, there is a limitation on the compressionality of the system that arises from the energy equation, namely 
\begin{gather}
    1 \leq r \leq \frac{\gamma +1}{\gamma -1}.
\end{gather}
For $\gamma =5/3$, this leads to a maximum compression ratio of $r=4$.

\subsection{Collisional ionisation and recombination (no ionisation potential) (IR model)} \label{sec:anIR}

When the two-fluid equations are used with ionisation and recombination (IR) in the absence of ionisation potential terms, one can again find an analytical solution for the jumps across the shock. For the deHoffmann-Teller shock frame, we set the time derivatives to be zero and assume the electric field is zero across the shock. 
\begin{gather}
\nabla \cdot (\rho _{\text{n}} \textbf{v}_{\text{n}})= S_{mass}, \\
\nabla \cdot (\rho _{\text{n}} \textbf{v}_{\text{n}} \textbf{v}_{\text{n}} + P_{\text{n}} \textbf{I}) = S_{mom}, \\
\nabla \cdot \left[\textbf{v}_{\text{n}} (e_{\text{n}} +P_{\text{n}}) \right] = S_{eng}, \\
\nabla \cdot (\rho_{\text{p}} \textbf{v}_{\text{p}}) = - S_{mass},\\
\nabla \cdot \left( \rho_{\text{p}} \textbf{v}_{\text{p}} \textbf{v}_{\text{p}} + P_{\text{p}} \textbf{I} - \textbf{B B} + \frac{\textbf{B}^2}{2} \textbf{I} \right) = - S_{mom}, \\
\nabla \cdot \left[ \textbf{v}_{\text{p}} ( e_{\text{p}} + P_{\text{p}}) -  (\textbf{v}_p \times \textbf{B}) \times \textbf{B} \right] =  -S_{eng}, \\
\nabla \times (\textbf{v}_{\text{p}} \times \textbf{B}) = 0, \\
S_{mass}=\Gamma _{rec} \rho _p - \Gamma _{ion} \rho _n, \label{eqn:smass}\\
S_{mom}=-\alpha _c \rho_{\text{n}} \rho_{\text{p}} (\textbf{v}_{\text{n}}-\textbf{v}_{\text{p}}) + \Gamma _{rec} \rho _p \textbf{v}_{p} - \Gamma _{ion} \rho_n \textbf{v}_n, \label{eqn:smom}\\
S_{eng} =-\alpha _c \rho _{\text{n}} \rho _{\text{p}} \left[ \frac{1}{2} (\textbf{v}_{\text{n}} ^2 - \textbf{v}_{\text{p}} ^2)+ \frac{3}{2} \left(\frac{P_n}{\rho_n}-\frac{1}{2}\frac{P_p}{\rho_p}\right) \right] \nonumber \\ \hspace{1.0cm}+ \frac{1}{2} \left( \Gamma _{rec} \rho _p \textbf{v}_p ^2 - \Gamma _{ion} \rho _n \textbf{v}_n ^2 \right) \nonumber \\ \hspace{1.0cm}  +\frac{1}{ (\gamma-1)} \left( \frac{1}{2} \Gamma _{rec} P_p -\Gamma _{ion} P_n \right). \label{eqn:seng} 
\end{gather}
However these equations have a non-zero right hand side due to the source terms and therefore cannot be integrated in their current form.

Since the IR equations are conservative, the source terms are equal and opposite for the plasma and neutral equations, i.e., momentum/energy/mass lost/gained by the neutrals is gained/lost by the plasma. As such, the equations can be added together to remove the source terms and integrated. The partial densities and pressures can also be expressed in terms of the neutral fraction ($\xi_n$), bulk density ($\rho_B$) and bulk pressure ($P_B$).

\begin{gather}
    \left[\xi_n \rho_{B} v_{nx} +(1-\xi_n) \rho_{B} v_{px}  \right]^u _d = 0,  \\
    \left[\rho_{B} v_{nx}^2 + \frac{\xi _n}{ 2-\xi_n} P_{B} +(1-\xi_n)\rho _{B} v_{px}^2 \right. \nonumber \\ \hspace{0.5cm} \left. +\frac{2(1-\xi _n)}{2-\xi_n} P_{B} +\frac{B_y^2}{2} \right]^u _d = 0, \\
    \left[\xi_n \rho_{B} v_{nx} v_{ny}  + (1-\xi_n) \rho _{B} v_{px} v_{py} -B_x B_y \right]^u _d = 0, \\
    \left[ v_{nx} \left( \frac{\gamma}{\gamma -1} \frac{\xi _n}{ 2-\xi_n} P_B + \frac{1}{2} \xi _n \rho _B v_n^2 \right) \right. \nonumber \\ \hspace{0.5cm} \left. +v_{px} \left( \frac{\gamma}{\gamma -1} \frac{2(1-\xi _n)}{ 2-\xi_n} P_B + \frac{1}{2} (1-\xi _n) \rho _B v_p^2 \right) \right]^u _d =0, \\
    \left[B_x \right]^u _d = 0,\\
    \left[v_{px} B_{y} -v_{py} B_x   \right]^u _d = 0,
\end{gather}
for upstream (superscript $u$) and downstream (superscript $d$) states. 

To simplify the equations further one can make some assumptions regarding the medium far away from the shock. Sufficiently upstream and downstream of the shock, one would expect that the drift velocity is zero, i.e., $v_{nx}=v_{px}$, $v_{ny}=v_{py}$. This greatly simplifies the equations:
\begin{gather}
    \left[\rho_B v_x  \right]^u _d = 0,  \\
    \left[\rho_B v_x^2 +P_B +\frac{B_y^2}{2} \right]^u _d = 0, \\
    \left[\rho_B v_x v_y -B_x B_y \right]^u _d = 0, \\
    \left[ v_{x} \left( \frac{\gamma}{\gamma -1} P_B + \frac{1}{2} \rho _B v^2 \right) \right]^u _d =0,\\
    \left[B_x \right]^u _d = 0, \\
    \left[v_x B_y -v_y B_x   \right]^u _d = 0. 
\end{gather}
This set of equations is identical to the MHD shock equations meaning that the analytic solution is identical to the MHD solution. This only holds sufficiently upstream and downstream of the shock since two-fluid effects lead to a finite shock width and substructure \citep{Snow2019}. 

\subsection{Collisional ionisation, recombination and ionisation potential (IRIP model)} \label{sec:anIRIP}

\begin{figure}
    \centering
\includegraphics[width=0.99\linewidth,clip=true,trim=1.0cm 8.2cm 1.2cm 8.3cm]{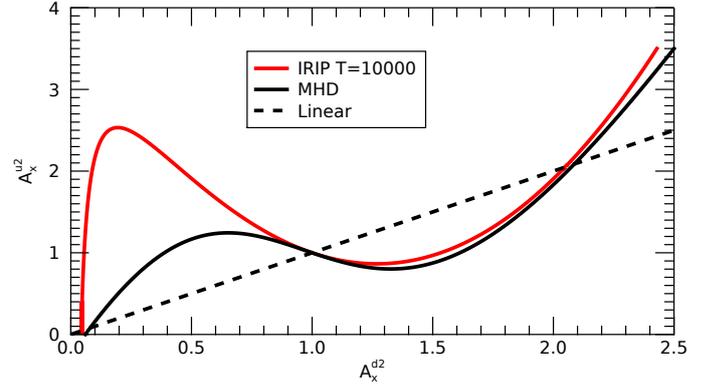}
    \caption{Solutions to the shock jump equations for the IRIP (red) and MHD (black) equations relating the upstream ($^u$) and downstream ($^d$) Alfv\'en Mach numbers $A_x$. The trivial solution ($A_x^{d}=A_x^{u}$) exists for both sets of equations. The parameters used for this plot are $T_0 = 10000$ K, $\beta = 0.1$, $\theta =\pi/4$ and $\gamma =5/3$.}
    \label{fig:tjumpex}
\end{figure}

Including ionisation potential and arbitrary heating for the IRIP model introduces two new terms into the plasma energy equation:
\begin{gather}
\nabla \cdot (\rho _{\text{n}} \textbf{v}_{\text{n}})= S_{mass}, \\
\nabla \cdot (\rho _{\text{n}} \textbf{v}_{\text{n}} \textbf{v}_{\text{n}} + P_{\text{n}} \textbf{I}) = S_{mom}, \\
\nabla \cdot \left[\textbf{v}_{\text{n}} (e_{\text{n}} +P_{\text{n}}) \right] = S_{eng}, \\
\nabla \cdot (\rho_{\text{p}} \textbf{v}_{\text{p}}) = - S_{mass},\\
\nabla \cdot \left( \rho_{\text{p}} \textbf{v}_{\text{p}} \textbf{v}_{\text{p}} + P_{\text{p}} \textbf{I} - \textbf{B B} + \frac{\textbf{B}^2}{2} \textbf{I} \right) = - S_{mom}, \\
\nabla \cdot \left[ \textbf{v}_{\text{p}} ( e_{\text{p}} + P_{\text{p}}) -  (\textbf{v}_p \times \textbf{B}) \times \textbf{B} \right] =  -S_{eng} -\phi _I +A_{heat}, \\
\nabla \times \left(\textbf{v}_{\text{p}} \times \textbf{B} \right) = 0, \\
\end{gather}
with $S_{mass},S_{mom},S_{eng}$ given by Equations (\ref{eqn:smass}-\ref{eqn:seng}).
Since the ionisation potential $\phi_I$ and the arbitrary heating $A_{heat}$ only exist in the plasma energy equation, 
adding together the equations for neutral and plasma species results in:
\begin{gather}
    \nabla \cdot (\rho _n \textbf{v}_n +\rho _p \textbf{v}_n)= 0, \\
    \nabla \cdot (\rho _n \textbf{v}_n \textbf{v}_n + P_n \textbf{I} +\rho _p \textbf{v}_p \textbf{v}_p + P_p \textbf{I}) = 0, \\
    \nabla \cdot \left[\textbf{v}_n (e_n +P_n) +\textbf{v}_p (e_p +P_p) -  (\textbf{v}_p \times \textbf{B}) \times \textbf{B} \right] \nonumber \\ \hspace{2cm} = -\phi_I + A_{heat}, \label{eqn:iripen} \\
    \nabla \times \left(\textbf{v}_p \times \textbf{B} \right) = 0, \\
    \nabla \cdot \textbf{B} = 0.
\end{gather}
Now it is fairly clear that all except the energy equation (\ref{eqn:iripen}) can be trivially integrated across the shock. Since the energy equation is non-conservative, we cannot use this equation to derive any shock jump conditions. A semi-analytical solution can still be found by replacing the energy equation with the relation that the ionisation potential and arbitrary heating terms must balance sufficiently upstream and downstream of the shock, Equation (\ref{eqn:irbulk}). 
The assumption of zero drift sufficiently upstream and downstream is also applied to get a solution for the bulk fluid. Our shock jump equation for the IRIP model are then:
\begin{gather}
    \left[\rho_B v_x  \right]^u _d = 0, \label{eqn:iripjump1} \\
    \left[\rho_B v_x^2 +P_B +\frac{B_y^2}{2} \right]^u _d = 0, \\
    \left[\rho_B v_x v_y -B_x B_y \right]^u _d = 0, \\
    \left[B_x \right]^u _d = 0, \\
    \left[v_x B_y -v_y B_x   \right]^u _d = 0, \label{eqn:iripjump2} \\
    \left[ F(T) \xi_i^2 \rho_B^2 \right]^u _d=0.
\end{gather}
The ideal gas law can be used to relate the temperature and pressure and close the system.
Therefore, all possible shock transitions can be obtained by solving the following equations:
\begin{gather}
    \frac{T^d}{T^u} = \frac{A_x^{d2}}{A_x^{u2}} \left[1 + \frac{2}{\beta (1+ \tan ^2 (\theta))} \times \nonumber \right. \\ \left. \hspace{0.7cm} \left[ A_x^{u2}-A_x^{d2} + \frac{\tan ^2 (\theta)}{2} \left(1 - \left(\frac{A_x^{u2}-1}{A_x^{d2}-1}\right)^2\right)  \right] \right], \label{eqn:irtjump} \\
    \frac{F(T ^d)}{F(T^u)} \left( \frac{F(T^u)/G(T^u) +1}{F(T^d)/G(T^d) +1} \right)^2= \frac{A_x^{d4}}{A_x^{u4}}. \label{eqn:ftjump}
\end{gather}
A derivation of these equations is included in Appendix \ref{app:iripjump}. 
Equation (\ref{eqn:irtjump}) gives a temperature jump across the shock (using the ideal gas law and Equations \ref{eqn:iripjump1}-\ref{eqn:iripjump2}) and can be solved numerically in conjunction with Equation \ref{eqn:ftjump} to get the possible shock transitions for a given upstream plasma-$\beta$, angle of magnetic field, and reference temperature. A solution for $T^u=10000$ K is given in Figure \ref{fig:tjumpex}. The trivial solution ($A^u=A^d$) still satisfies the IRIP jump conditions and has three intersections with the non-trivial solution. 

Both the IRIP and MHD models have an intersection when the upstream and downstream Alfv\'en Mach numbers are unity which is the rotational discontinuity. For $A^{d}<1$ there are slow ($A^{u}<1$) and intermediate ($A^{d}>1,A^{u}<1$) shock solutions. Interestingly, these solution imply much greater compression than the MHD solution since the upstream Alfv\'en Mach number for the IRIP case is much larger relative to the downstream Alfv\'en Mach number. Unlike the MHD model, there is no restriction on the compressionality of the system.

In this paper, the focus is on a switch-off slow-mode shock which occurs when the upstream Alfv\'en Mach number is unity. From Figure \ref{fig:tjumpex} it can easily be seen that the switch-off shock in the IRIP model will have a much smaller downstream shock velocity than in the MHD case. It will also have a much greater compressional ratio and will have a decrease in temperature across the shock.




\begin{figure*}
    \centering
\includegraphics[width=0.95\linewidth,clip=true,trim=1.5cm 7.8cm 1.5cm 7.8cm]{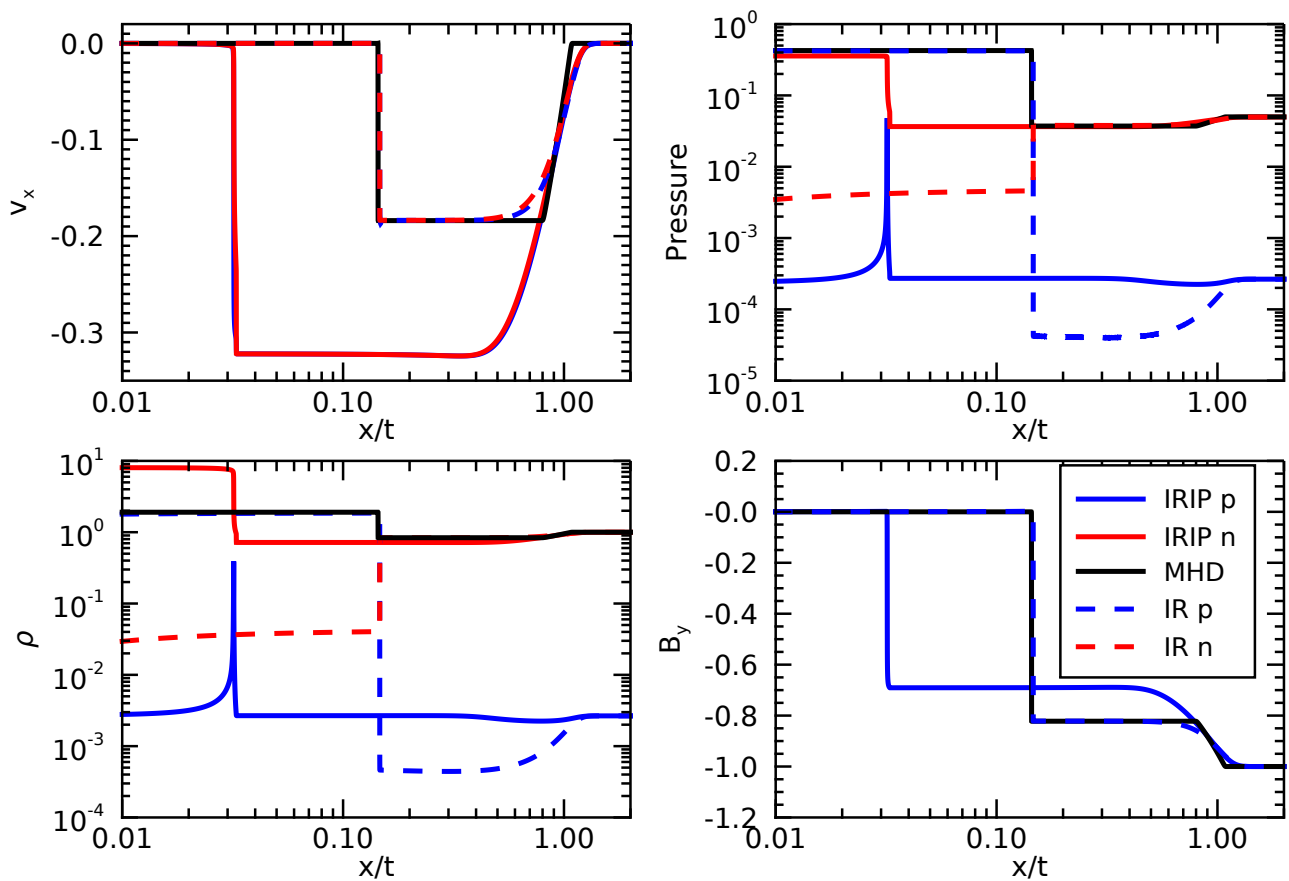}
    \caption{Context figure for the equilibrium state in the MHD (black), IRIP (solid) and IR (dashed) cases. The blue (red) line is for the plasma (neutral) species for the two-fluid cases.}
    \label{fig:radfull}
\end{figure*}

\section{Numerical model}

Numerical simulations are performed using the (P\underline{I}P) code \citep{Hillier2016} that solves for two fluids (neutral, and ion+electron) in normalised units. The (P\underline{I}P) code has been developed since earlier editions and now features terms to account for ionisation, recombination and ionisation potential energy. Specifically, Equations (\ref{eqn:neutral1}-\ref{eqn:plasma2}) are solved numerically using a first order HLLD scheme. The source terms for collisions and ionisation/recombination are integrated explicitly. The system is resolved by 256000 grid cells.

\subsection{Initial conditions}

The initial conditions are designed to mimic a slow-mode shock produced by fast magnetic reconnection. Specifically, our initial conditions consist of a thermal equilibrium and a discontinuity in magnetic field across the origin:

\begin{gather}
B_x = 0.1 \\
B_y = -1.0 (x>0), 1.0 (x<0) \\
\rho _{\text{n}} = \xi _{\text{n}} \rho _{\text{t}} \\
\rho _{\text{p}} = \xi _i \rho _{\text{t}} \textbf{=} (1- \xi _{\text{n}}) \rho _{\text{t}} \\
P_{\text{n}} = \frac{\xi _{\text{n}}}{\xi_{\text{n}} + 2 \xi _i} P_{\text{t}} =  \frac{\xi _{\text{n}}}{\xi_{\text{n}} + 2 \xi _i} \beta \frac{B_0 ^2}{2} \\
P_{\text{p}} = \frac{2 \xi _i}{\xi_{\text{n}} + 2 \xi _i} P_{\text{t}} =  \frac{2 \xi _i}{\xi_{\text{n}} + 2 \xi _i} \beta \frac{B_0 ^2}{2}
\end{gather}
where $\xi _n$ and $\xi_i$ are the neutral and ion fractions respectively. These initial conditions have been studied in the absence of ionisation and recombination in \cite{Hillier2016,Snow2019}.

In normalised form, the equilibrium neutral fraction depends on the reference temperature $T_0$. Taking the continuity of mass (Equation \ref{eqn:neutral1}) and rearranging, we get $\Gamma _{rec}/\Gamma _{ion} = \rho _n/ \rho _p$. Therefore the equilibrium neutral fraction $\xi_n$ can be calculated using 
\begin{gather}
\xi _n = \frac{\Gamma _{rec}/\Gamma _{ion}}{(\Gamma _{rec}/\Gamma_{ion} +1)}    
\end{gather}
The initial neutral fraction is controlled by varying the reference temperature $T_0$.

We consider three different cases in this work. Firstly, for reference, the MHD model is considered where the plasma is fully-ionised. Secondly, we have the IRIP model for a two-fluid system where ionisation, recombination, ionisation potential and arbitrary heating is included. Finally, again for reference, an IR model is used which is multi-fluid with ionisation and recombination but neglecting the ionisation potential and heating terms. The main focus of this paper is the IRIP model.


\section{Results} \label{sec:results}

For the initial simulations we use $T_0=10000$ K which leads to an equilibrium neutral fraction of $\xi_n \approx 0.99734$. We set the recombination rate to be $\Gamma _{rec} = 10^{-3}$ and the collisional frequency is set to unity. Physically, this means that collisions occur on a timescale of unity, and recombination occurs on a timescale of 1000. The ionisation timescale is calculated from the recombination timescale and the normalisation temperature. Using $T_0=10000$ K, the initial ionisation rate is $\Gamma _{ion} = 2.66 \times 10^ {-6}$. The recombination and ionisation rates depend on the instantaneous temperature of the system which varies in time and hence the rates also vary. 

\subsection{Equilibrium solution ($t=40000$)}

After 40000 collisional times, the system is sufficiently steady. The rarefaction wave and shock have separated, and variables between these features have very small gradients. As such, an (approximate) equilibrium state has been reached. 
Figure \ref{fig:radfull} shows this pseduo-steady state for the IRIP model compared to the MHD solution. For comparison, an IR simulation is also presented which does not include the ionisation potential and heating terms.

The first result to state is that the IRIP result is very different to the MHD and IR solutions. This is not unexpected for the reasons given in previous sections, namely that the energy equation is non-conserving, equilibrium only exists at certain densities, and we expect greater compression across the shock. In particular, the inflow to the shock front is much larger in the IRIP case, and the propagation speed of the IRIP shock is much lower than the MHD case.

\subsubsection{Inflow}

\begin{figure*}
    \centering
\includegraphics[width=0.95\linewidth,clip=true,trim=0.9cm 7.8cm 1.5cm 7.8cm]{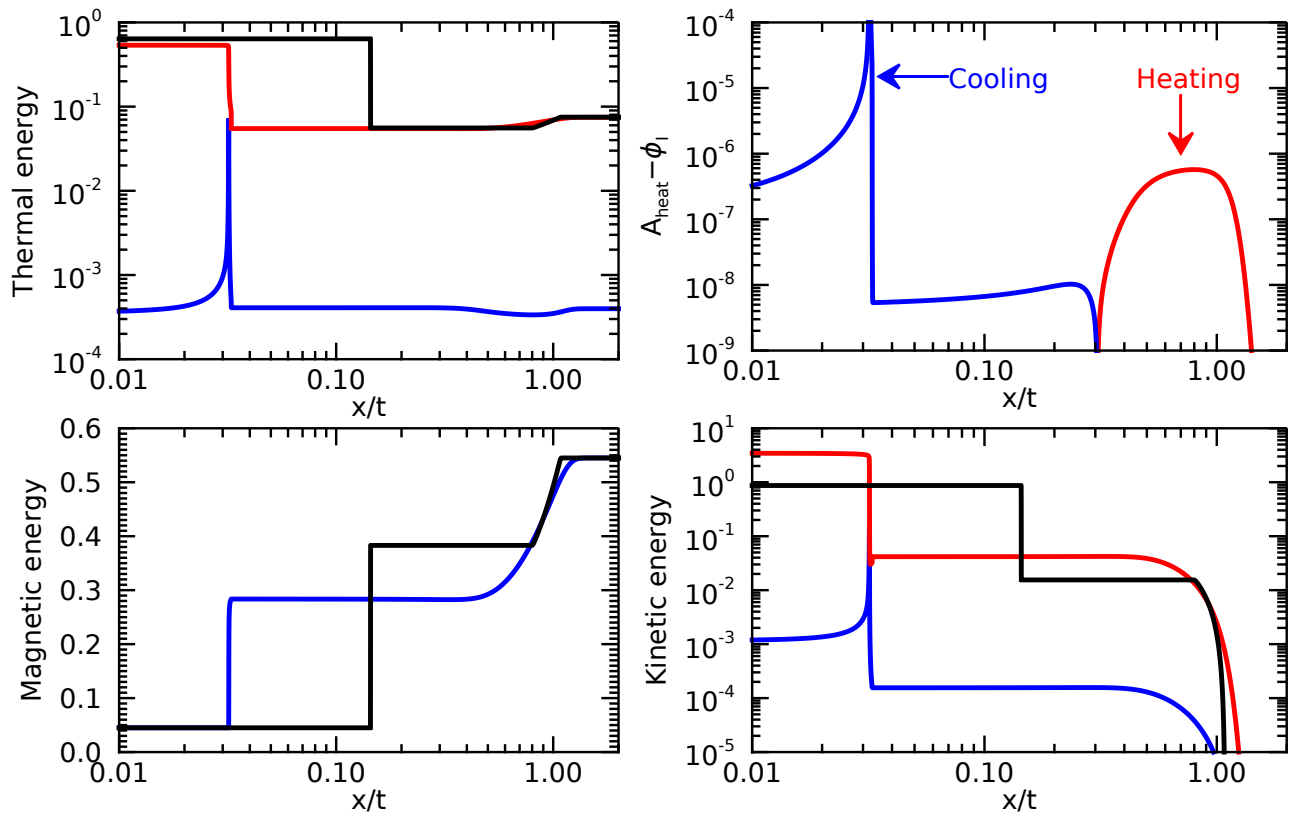}
    \caption{Energies across the system for the MHD (black) and IRIP cases (red neutrals, blue plasma). The quantity $A_{heat}- \phi _I$ shows the balance between the heating and loss terms in the energy equation where red denotes net energy addition (heating) and blue is net energy loss (cooling).}
    \label{fig:rareeng}
\end{figure*}

\begin{figure*}
    \centering
\includegraphics[width=0.95\linewidth,clip=true,trim=0.9cm 7.8cm 1.5cm 7.8cm]{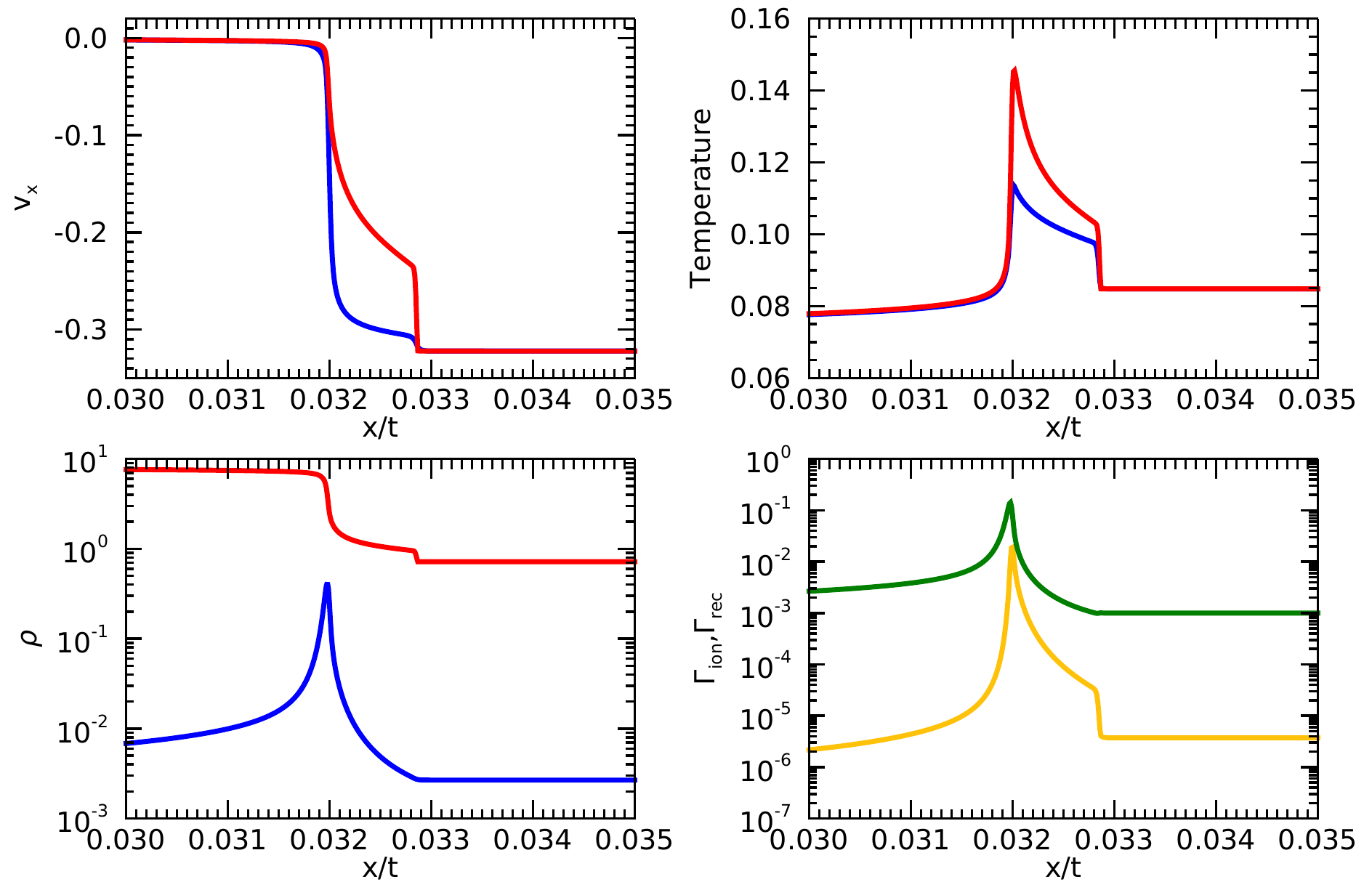}
    \caption{Close-up of the slow mode shock for the IRIP model showing $v_x$ velocity (top left), temperature (top right) and density (lower left) for plasma (blue) and neutral (red) species. The lower right panel shows the ionisation (orange) and recombination (green) rates.}
    \label{fig:radslow}
\end{figure*}

To understand the inflow to the shock, the physics of the rarefaction wave must be understood. Figure \ref{fig:rareeng} shows the energy terms across the system. For the MHD rarefaction wave, magnetic and thermal energy are converted to kinetic energy driving an inflow towards the slow-mode shock. 


For the IRIP case, there is a similar thermal energy reduction, and a much larger loss of magnetic energy, compared to MHD, see Figure \ref{fig:rareeng}. The thermal energy reduction is mostly in the neutral species which constitutes the majority of the bulk fluid. The rarefaction wave is expansive, meaning that there is a decrease in total density across the rarefaction wave. The density reduction is in the neutral species and plasma density is roughly the same in the static medium and inflow regions, see Figure \ref{fig:radfull}.



In the IRIP case, the total density decrease across the rarefaction wave results in a slight increase in temperature due to the requirement that for a steady (equilibrium) solution to exist
\begin{gather}
    G(T)\rho _n \rho_p =F(T) \rho_p^2=\Gamma_{ion}(t=0)\rho_n(t=0)=\mbox{const}.
\end{gather}
The decreased total density necessitates an increased ionisation rate which can only be obtained through a temperature increase of the plasma species (since both the ionisation and recombination rates rely on the plasma temperature). The temperature increase is provided through the arbitrary heating term, which, as the system undergoes expansion, is larger than the energy lost through ionisation potential in the rarefaction wave. 



The leading edge of the rarefaction wave propagates at the same speed in the MHD and IRIP models. However, the IRIP rarefaction wave has a far greater expansion than the MHD case. 
The rarefaction wave in all simulations drives velocity towards the slow-mode shock at the Alfv\'en speed. 

Across the rarefaction wave, the IR jumps in bulk quantities match the MHD solution (see Figure \ref{fig:radfull}). This matches the theoretical result from Section \ref{sec:anIR} that, since the IR equation reduce to the MHD equations sufficiently upstream and downstream of a feature, the MHD result should be obtained. 



\subsubsection{Slow-mode shock} \label{sec:slowshock}

\begin{figure}
    \centering
\includegraphics[width=0.95\linewidth,clip=true,trim=0.9cm 7.8cm 1.5cm 7.8cm]{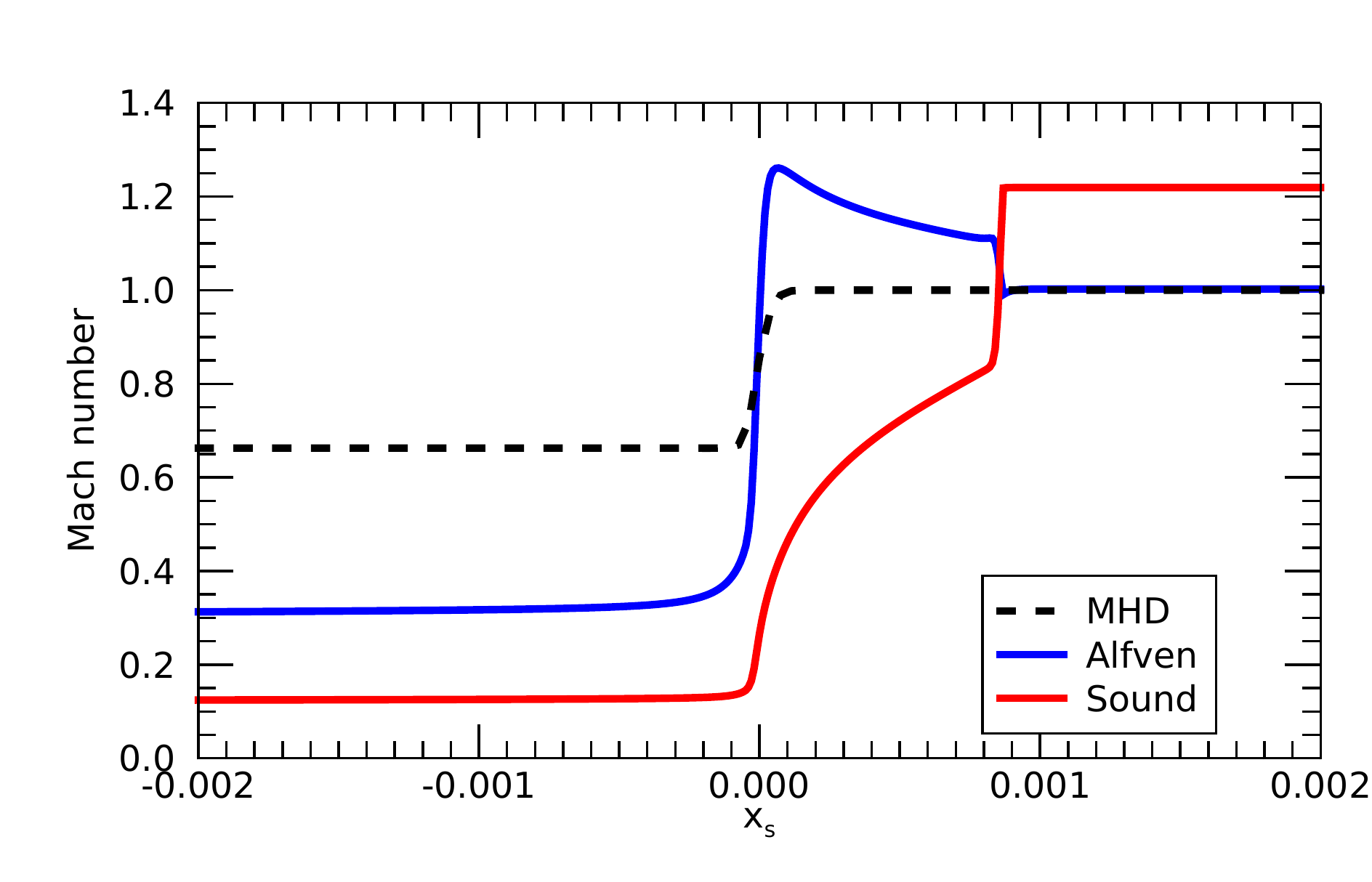}
    \caption{Alfven Mach numbers for the MHD (black) and plasma (blue) based on the bulk density. The red line shows the neutral sonic Mach number.}
    \label{fig:mach}
\end{figure}

\begin{figure*}
    \centering
\includegraphics[width=0.95\linewidth,clip=true,trim=0.9cm 7.8cm 0.5cm 7.8cm]{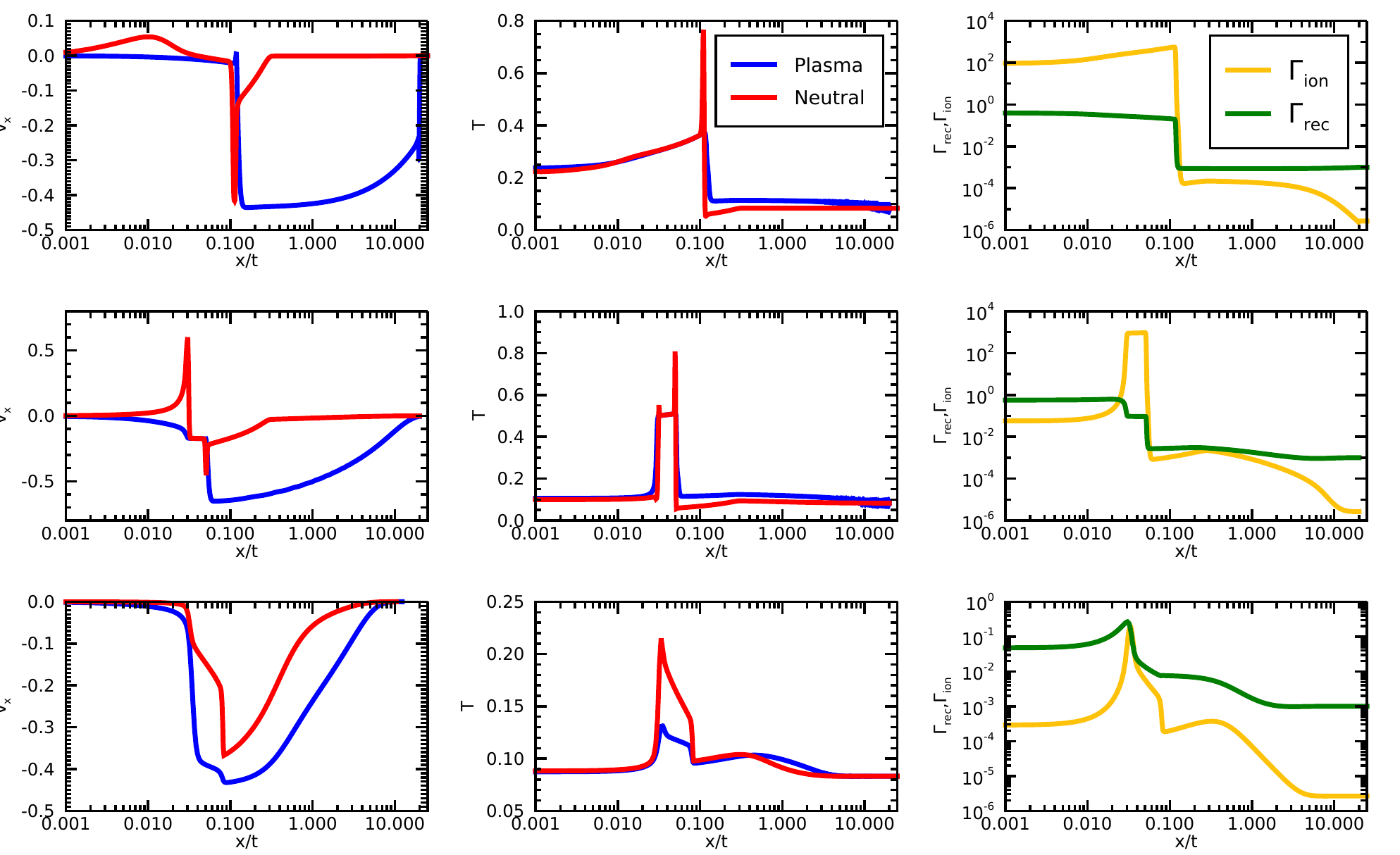}
    \caption{Time series for the IRIP case after 1 (top), 10 (middle) and 100 (lower) collisional times showing the $v_x$ velocity (left), temperature (centre), and ionisation and recombination rates (right).}
    \label{fig:timeseries}
\end{figure*}

The inflow conditions for the slow-mode shock in the IRIP case are different to the MHD case, as such the jump conditions are expected to be different. Figure \ref{fig:radslow} shows the velocity (top left), temperature (top right), density (lower left), and ionisation and recombination rates (lower right) across the slow mode shock. Since the system has been evolved for 40000 collisional times the slow-mode shock is fairly localised on the $x/t$ axis but is well resolved by our grid resolution. 

The rarefaction wave drives inflow towards the slow-mode shock. In the IRIP case, the $v_x$ velocity is far greater than in the MHD case, however, in the shock frame, both systems have an inflow Alfv\'en Mach number of one, see Figure \ref{fig:mach}. The specific shock type here is a switch-off slow-mode shock where the perpendicular magnetic field reduced to zero across the shock. Switch-off shocks are the most extreme slow-mode shocks and occur when the inflow Alfv\'en Mach number is unity.  

The velocity structure across the shock (Figure \ref{fig:radslow}) displays similar structures to the PIP case in \cite{Hillier2016,Snow2019} namely that the neutral species shocks first, followed by a gradual coupling to the plasma resulting in a finite width to the slow-mode shock. Two-fluid effects result in a finite shock width where the species separate and interesting physics occurs.

In the absence of ionisation and recombination, a key feature of two-fluid interaction in shocks is an overshoot of the neutral velocity \citep{Hillier2016,Snow2019}. The large drift between the ions and neutrals inside the shock results in frictional heating leading to a Sedov-Taylor-like expansion of the neutral species.
The neutral overshoot is not present in the IRIP model presented here (see Figure \ref{fig:radslow}). The initial discontinuity drives the plasma only, which then couples to the neutral species. At early times in the simulation ($t=10$) an overshoot in the neutrals exists but ionisation prevents it from being a sustained feature of the system (the time evolution of the system is discussed further in Section \ref{sec:earlyt}).    


At the leading edge of the shock, a sonic shock occurs in the neutral species, see Figure \ref{fig:mach}. The sonic shock features a temperature increase (due to adiabatic heating) in the neutral species. Collisional coupling transfers this heat from the neutrals to the plasma and hence leads to a local enhancement of ionisation. Behind this, the plasma produces a slow-mode shock, with another localised temperature increase and ionisation rate.

The post-shock region has a lower temperature than the pre-shock region. This is related to the relation:
\begin{gather}
    G(T)\rho _n \rho_p =F(T) \rho_p^2=\Gamma_{ion}(t=0)\rho_n(t=0).
\end{gather}
Both the slow-mode and the sonic shock that forms within it are compressional, thus the increase in density leads to a decrease in temperature from the ionisation potential energy losses. An interesting corollary of the temperature reduction across the shocks is that observational signatures of chromospheric shocks may manifest as a decrease in intensity due to the temperature drop. Note that the slow-mode shock features an increase in both density and pressure across the shock.


The Mach numbers for the system in the shock frame are shown in Figure \ref{fig:mach}. The MHD case is a slow-mode switch-off shock, which is the strongest possible slow-mode shock where the inflow velocity is the Alfv\'en speed. In the IRIP case, despite the very different inflow velocity and magnetic field, we still have an Alfv\'enic inflow and a super- to sub-slow transition across the shock indicating the same switch-off slow-mode shock as in the MHD case. In the two-fluid case, the species decouple and recouple around the shock front. At the leading edge of the shock the species separate and a sonic shock forms in the neutral species, as seen in \cite{Hillier2016}. Also, the plasma exceeds the Alfv\'en speed inside the finite width of the shock, resulting in an intermediate shock. The magnitude of the magnetic field reversal associated with the intermediate shock decreases with time due to the balance between the neutral pressure and the Lorentz force, as described in \cite{Snow2019}. After 40000 collisional times, the magnetic field reversal is very small.


Using the upstream properties from the numerical simulation in analytical solution, in inflow Alfv\'en Mach number pairs remarkably well (see Figure \ref{app:iripjump}). This gives us confidence in the applicability of the semi-analytical solution described in Section \ref{sec:anIRIP}. 

\subsection{Early times} \label{sec:earlyt}

Now that the equilibrium state is understood, the system can be considered on transient timescales. For this simulation, the initial recombination rate was set to $\Gamma _{rec}=10^{-3}$ of the collisional time. As such, one expects the system to reach a collisional equilibrium long before an ionisation equilirium is reached.

After 1 collisional time the neutrals have experienced very few collisions. The initial conditions feature a discontinuity of magnetic field across the origin, triggering a fast-mode wave and a slow-mode shock in the plasma species only. At such an early time, the fast-mode and slow-mode are still interacting and have not decoupled. The slow-mode shock features a temperature increase and hence increased local collisions so the neutral species has a slight response. The increased temperature boosts the ionisation and recombination rates by approximately nine and two orders of magnitude respectively. As such, these effects become important even on such small timescales. The large increase in ionisation rate results in large energy losses through the ionisation potential in these regions.

After 10 collisional times, the neutral species has a clear response to the plasma motion due to collisional coupling. The profiles are still very distinct and far from equilibrium state. The ionisation and recombination effects are starting to have an effect on the system here, as can be seen by the post-shock cooling in the plasma temperature.

By 100 collisional times, similar structure begins to exist in the plasma and neutral species. A sonic shock can be seen in the neutral species. The post shock temperature is very similar to the case after 40000 collisional times in Figure \ref{fig:radslow}. The pre-shock region is still far from equilibrium, with large drift velocities existing between the two species.

\subsection{Steady solution}

\begin{figure}
    \centering
\includegraphics[width=0.95\linewidth,clip=true,trim=0.9cm 7.8cm 1.5cm 7.8cm]{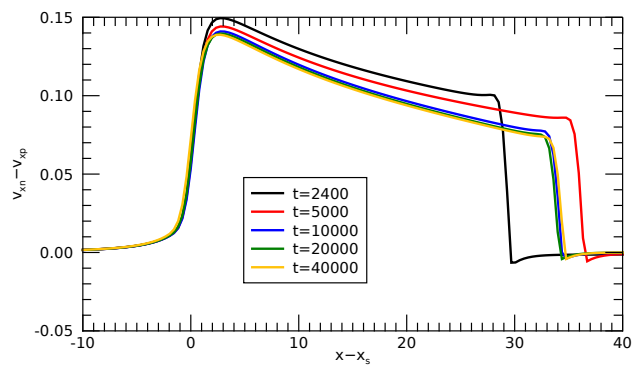}
    \caption{Drift velocity at different times for the IRIP simulation. The shocks are aligned using $x_s$ which is the location of the minimum gradient of the plasma velocity, i.e., the shock terminus. }
    \label{fig:sstest}
\end{figure}

As time advances, the shock tends towards a steady solution, where the finite-width of the shock is determined by the diffusive terms of the system. Figure \ref{fig:sstest} shows the drift velocity throughout the slow-mode shock at different times, and shock width can be determined by the departure from zero drift (i.e., coupled fluid). 
The peak drift velocity is approximately 15\% of the bulk Alfv\'en speed.
After 10000 collisional times, the shock width changes very little and the system can be considered to be a steady solution. The finite width of the shock is determined based on the physical diffusion mechanisms, here the ion-neutral collisions and ionisation, recombination and ionisation potential. 10000 collisional times corresponds to 10 recombination times based on the background recombination rate of $\tau_{IR} = 10^{-3}$. However, inside the shock the local temperature increases result in a recombination rate a few orders of magnitude higher than the background value, see Figure \ref{fig:radslow}. Hence recombination (and ionisation) occurs on a far faster timescale inside and around the shock than in implied by background quantities. Taking the peak recombination rate inside the shock of $~0.1$, we can estimate that 10000 collisional times corresponds to approximately 1000 recombination times. As such, the system has had plenty of time to reach ionisation/recombination equilibrium inside the shock.     

The finite width of the shock is much smaller in the IRIP model than when ionisation and recombination is neglected. In the IRIP simulation, the finite width is approximately $L=35$. For a simulation using the same initial neutral fraction but without ionisation and recombination, the shock width is approximately $L=414$. The smaller finite width can be explained by the increased ion-fraction inside the shock shown in Figure \ref{fig:radslow} \citep[this was shown in][where a higher ion-fraction leads to a smaller shock]{Hillier2016}. A consequence of the reduced shock width is that a package of fluid moving though the shock will lose less energy than it would through a wider shock. As such, cooling inside the shock becomes less important for a narrower shock. The post-shock cooling is still very important in removing energy from the system.   


\section{Parameter study}
\subsection{Recombination timescale}


\begin{figure}
    \centering
\includegraphics[width=0.95\linewidth,clip=true,trim=0.95cm 8.3cm 10.5cm 7.8cm]{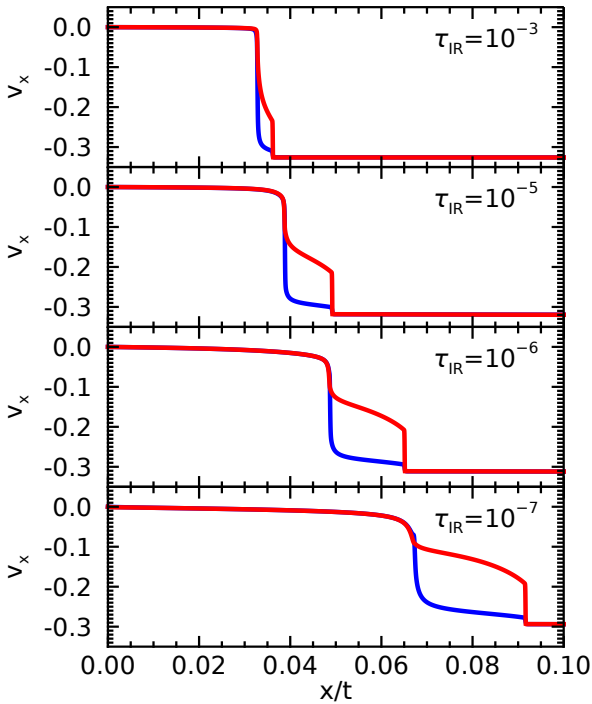}
    \caption{Plasma (blue) and neutral (red) $v_x$ velocities for recombination time scales of $10^{-3}$, $10^{-5}$, $10^{-6}$, $10^{-7}$ from top to bottom.}
    \label{fig:shockwidthvx}
\end{figure}

\begin{figure}
    \centering
    \includegraphics[width=0.95\linewidth,clip=true,trim=0.9cm 7.8cm 1.2cm 7.8cm]{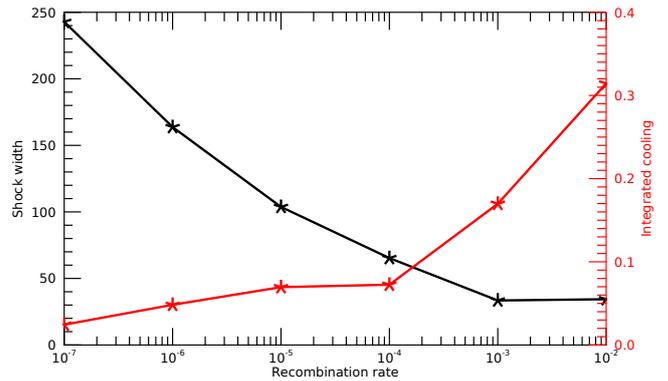}
    \caption{Finite width of the shock (black line) as a function of the initial recombination rates. Integrated cooling for a parcel of fluid travelling through the shock (red line).}
    \label{fig:shockwidthsw}
\end{figure}

\begin{figure}
    \centering
\includegraphics[width=0.95\linewidth,clip=true,trim=0.9cm 7.8cm 1.5cm 7.8cm]{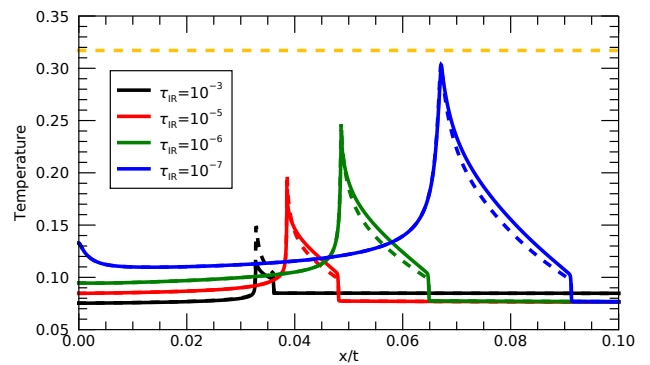}
    \caption{Plasma (solid) and neutral (dashed) temperatures for different initial recombination rates. The orange dashed line shows the downstream MHD temperature.}
    \label{fig:radt}
\end{figure}

\begin{figure}
    \centering
\includegraphics[width=0.95\linewidth,clip=true,trim=0.7cm 7.8cm 1.5cm 7.8cm]{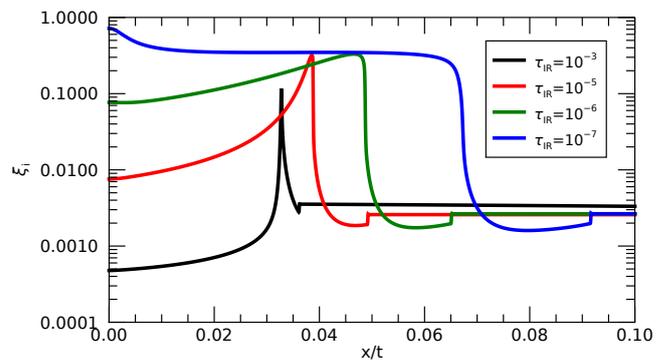}
    \caption{Ionisation fraction across the shock for different recombination rates.}
    \label{fig:shockwidthxi}
\end{figure}

In the previous simulations the initial recombination rate was scaled to $\tau_{IR}=10^{-3}$ of the collisional time. The initial ionisation rate is then a factor of the recombination time based on the reference temperature such that the system is in equilibrium. In this section we investigate changing the reference recombination rate $\tau_{IR}$, which also changes the ionisation rate. In Section \ref{sec:earlyt} it was found that the ionisation rate increases by approximately nine orders of magnitude at the slow-mode shock after 1 collisional time. Therefore, the rates are increased due to the initial temperature enhancements, resulting in the ionisation and recombination forces balancing before the collisional components equalise. As such, smaller reference recombination times are a primary focus in this section. All other system parameters are kept the same ($T_0=10000$ K, $\beta =0.1$).



Decreasing the recombination rate causes the shock to propagate more slowly and leads to a greater shock width, Figure \ref{fig:shockwidthvx}. Similar velocity structure exists in all simulations, i.e., a neutral shock at the leading edge, with a sharp jump in plasma velocity at the rear of the shock. In the shock frame, the inflow conditions for all simulations is the Alfv\'en speed, and as such all simulations can be classified as having a switch-off slow-mode shock. Plotting the shock width as a function of the initial recombination rate, Figure \ref{fig:shockwidthsw}, one can see the shock width decreases as the recombination rate increases. For $\tau _{IR} < 10^{-3}$ the system is still evolving and the shock is thinning. The $\tau _{IR}=10^{-2}$ case has roughly the same shock thickness as the $\tau_{IR}=10^{-3}$ case because the system has reached a steady state.

The initial conditions produce a large temperature from the adiabatic heating of the slow-mode shock in the plasma \citep[see][]{Hillier2016}. The temperature increase boosts the ionisation rate, which in turn increases the energy lost through ionisation potential.
Hence decreasing the rate of recombination (and with it, ionisation) results in energy being removed from the system slower and the medium undergoes more gradual cooling, see Figure \ref{fig:radt}. All simulations have the same restraint that the ionisation potential losses and arbitrary heating terms must balance for equilibrium and hence as distance from the shock front tends to infinity, the upstream temperature should be the same across all simulations. It can be concluded that all simulations are in collisional equilibrium (i.e., $v_p=v_n$ sufficiently upstream/downstream) however they are not in ionisation-recombination equilibrium (i.e., $\Gamma_{ion}\rho_n \neq \Gamma_{rec}\rho_p$). The low ionisation rates for $\tau_{IR}=10^{-7}$ result in a very gradual cooling and one would expect the equilibrium to be obtained on far longer timescales than studied here (end time $t=10000$). The decrease in cooling also affects the shock temperature since energy is being removed at a much lower rate. For $\tau=10^{-7}$ the shock temperature approaches the downstream MHD temperature, see Figure \ref{fig:radt}. 

The interior velocity structure is similar for the different initial recombination rates (Figure \ref{fig:shockwidthvx}) however the neutral fraction is very different, as shown in Figure \ref{fig:shockwidthxi}. In all cases, the neutral shock at the leading edge results in a decrease in ionisation fraction due to the increased losses from the adiabatic heating. Following this, there is a gradual rise in the ionisation fraction throughout the shock, reaching an apex at the shock terminus. Post-shock, the neutral fraction is much larger for smaller $\tau_{IR}$ due to the decreased cooling rate for these simulations, compared to the reference value. The snapshots were taken after 10000 collisioal times however it is known that the $\tau _{IR}=10^{-3}$ case only just reaches a steady state around this time. One would expect that decreasing the recombination rate by an order of magnitude increases the time required for the steady solution by an order of magnitude. Therefore it would require $10^8$ collisional times for the $\tau _{IR}=10^{-7}$ case to reach its steady solution.

A parcel of fluid travelling through the shock experiences an energy loss due to the ionisation potential loss term being larger than the arbitrary heating. The integrated cooling across the finite width of the shock can be calculated using:
\begin{gather}
    \mbox{Cooling} = - \Sigma _a ^b (A_{heat}-\phi _I)\frac{\Delta x}{v_{xp}-v_s},
\end{gather}
where $a,b$ are the physical limits of the shocks finite width, $\Delta x$ is the grid cell size, and $v_s$ is the shock velocity. The integrated cooling through the shock is plotted in Figure \ref{fig:shockwidthsw} for different initial recombination rates. Decreasing the initial recombination rate lowers the ionisation potential and hence results in less cooling through the shock, and a higher temperature is obtained for smaller reference recombination rates (Figure \ref{fig:radt}). The shock also increases in width for smaller reference recombination rates. As such, the cooling of a parcel of fluid travelling across the shock is not as severe as one may expect; reducing the reference recombination time from $10^{-2}$ to $10^{-7}$ only reduces the cooling across the shock by one order of magnitude.   

To put this parameter study in the context of the solar atmosphere, in dimensional units, ionisation/recombination rates are typically on the order of $10^{-3}-10^{-5}$ per second and are locally enhanced in shocks \citep{Carlsson2002}. For a VALC chromosphere, the ion-neutral ($\nu_{in}$) and neutral-ion ($\nu_{ni}$) collisional frequencies are approximately in the ranges of $10^3 < \nu_{in} < 10^6$ and $10^{-1} < \nu_{ni} < 10^2$ per second varying with height due to stratification \citep[from Figure 8 in][]{Popescu2019}. We therefore expect the ionisation/recombination rates to be between 2 and 7 orders of magnitude smaller than the neutral-ion collisional frequency. As such, this parameter study covers the expected range of recombination time scales expected in the lower solar atmosphere.

\subsection{Reference temperature}


\begin{figure}
    \centering
\includegraphics[width=0.95\linewidth,clip=true,trim=1.1cm 7.8cm 1.5cm 7.8cm]{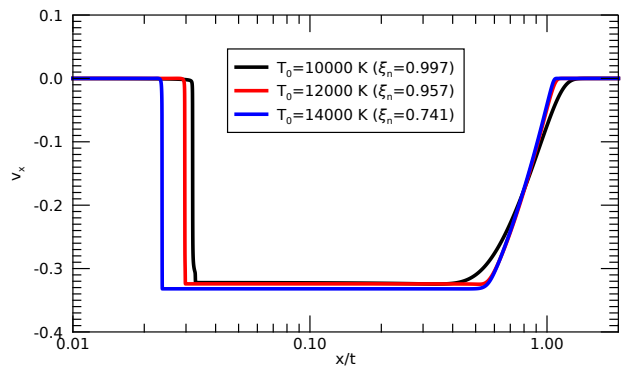}
    \caption{$v_x$ plasma velocity for different reference temperatures showing the equilibrium state after 40000 collisional times.}
    \label{fig:reftemp}
\end{figure}

The reference temperature is a fairly important parameter since is governs the initial neutral fraction and relative ionisation and recombination rates. Previously, the temperature was set to $T_0=10000$ K resulting in a predominantly neutral medium with a neutral fraction of $\xi_n\approx 0.99734$. In this section, the reference temperature is modified resulting in different neutral fractions, as shown in Figure \ref{fig:reftemp}. Other initial system parameters are kept as plasma-$\beta=0.1$ and recombination rate as $10^{-3}$.

The system studied produces a rarefaction wave that drives inflow towards a switch-off slow-mode shock at the Alfv\'en speed. At the equilibrium state, both the ionisation potential and heating terms must balance resulting in unique solutions for the density jump. All solutions show a similar equilibrium state to the one discussed in detail in Section \ref{sec:results}, see Figure \ref{fig:reftemp}. Increasing the reference temperature results in a slower propagating slow-mode shock, which is also predicted by the semi-analytical solution. Other than that, there are relatively few differences for different reference temperatures. Note that the simulations performed are in the partially ionised regime. As one tends towards the fully-ionised solution (i.e., MHD), one would expect greater differences to occur.  


\section{Discussion}
In the interstellar medium, molecules were discovered downstream of shocks that should have been disassociated due to the MHD predicted temperature inside the shock. However, including radiative losses in the system reduces the maximum obtained temperature in the shock, allowing these molecules to survive into the downstream medium \citep{Draine1993}. 
The results presented in this paper show a similar effect where there is a reduction in the maximum temperature obtained in the shock due to radiative losses (see Figure \ref{fig:radt}). A parcel of fluid moving through the shock experiences significant cooling and therefore lower temperature spectral lines may be present after the shock, compared with an MHD prediction. This result may be relevant to the IRIS result of \cite{depontieu2015} where the non-thermal line broadening in shocks was found to be greater than expected due to non-equilibrium ionisation. Non-equilibrium ionisation does not explain the full extend of the observational non-thermal broadening and hence the two-fluid effects discussed in this paper may be a way of increasing the non-thermal line width. Further study is needed to investigate the non-thermal line broadening due to two-fluid effects.   

Within the finite-width of the shock, drift velocities reach up to $15\%$ of the bulk Alfv\'en speed, as shown in Figure \ref{fig:sstest}. In terms of dimensional units, for plasma-$\beta=0.1$, and a chromospheric sound speed of 8 kms$^{-1}$, this drift velocity equates to an Alfv\'en speed of approximately 20 kms$^{-1}$ and hence a drift velocity of approximately 3 kms$^{-1}$. This is fairly substantial and lends credibility to a previously predicted signatures of two-fluid effects in shocks, namely the electric field felt by the neutral species \citep{Anan2017,Snow2019}.

Including the ionisation, recombination and ionisation potential terms narrows the finite width of the shock. However, the finite-width is far from discontinuous. A previously proposed effect of two-fluid propagating shocks is a large finite-width due to ion-neutral drag \citep{Snow2020}. This remains a potential observable for two-fluid shocks despite the narrowing of the finite-width, due to the large drift velocities present in our simulations. 

\section{Conclusions}

In this paper we analyse the consequences of collisional ionisation, recombination and ionisation potential on a partially-ionised switch-off slow-mode shock. The ionisation potential term leads to a non-conservative energy equations that results in a equilibrium shock structure very different from the MHD case.

For an equilibrium to exist, the heating and cooling terms must balance. As such, a simple equation can be derived that must be satisfied for equilibrium that relates the upstream temperature and density to the arbitrary heating constant. In the limit of $T<20000$ K, the function is monotonically increasing with temperature, implying that a compressional shock results in upstream cooling, as opposed to the temperature increase obtained from MHD results. It also implies that for a given temperature, equilibrium only exists at specific density values.

Shock jump equations can be derived for the IRIP model (two-fluid equations with collsional ionisation, recombination, ionisation potential and arbitrary heating) and a semi-analytic solution for possible shock transitions can be derived. The analytic solution for the shock Alfv\'en Mach number pairs well with the numerical solution for the slow-mode switch-off shock studied here. An empirical form of ionisation and recombination was used in this paper however the methodology could easily be applied to more advanced models. 

The semi-analytic jump conditions derived state that for a compressible, steady-state, partially ionised, shock with ionisation potential included, the temperature should decrease across the shock, see Equation \ref{eqn:ircomp}. For an equilibrium to exist, the ionisation potential energy losses must balance the arbitrary heating term. In the partially ionised regime ($T<20000$ K), the equilibrium constraint necessitates that a steady-state compressible shock has a decrease in temperature across the shock. Our numerical simulation show this result, where sufficiently downstream of the shock, the temperature is lower than upstream, as shown by Figure \ref{fig:radslow}. This is contrary to the MHD and IR result where a compressible shock requires a temperature increase across the interface. Furthermore, including ionisation potential leads to cooling of the plasma within the finite-width of the shock and lowers the maximum temperature obtained, as shown in Figure \ref{fig:radt}. As such, including ionisation potential fundamentally changes the behaviour of shocks in partially ionised plasmas.

An intermediate transition exists within the finite shock-width, as seen in the two-fluid simulations of \cite{Snow2019} without ionisation and recombination. This is a promising result that provides further evidence that additional shock transitions may occur within two-fluid shocks in the solar atmosphere.

Similar results are obtained when the recombination rate or reference temperature are altered. Decreasing the recombination rate results in the system taking longer to reach a steady state. Increasing the reference temperature reduces the ionisation fraction and results in a slower propagating shock. The slower propagation can be predicted from the semi-analytical solution. 



In conclusion, non-conservative energy effects can have significant consequences for shocks in the solar chromosphere. In particular, for the equations studied here, a compressional shock will result in downstream cooling of the system if the medium is in the partially ionised regime. A semi-analytical solution for the equilibrium shock jumps can be derived which pairs well with the numerical simulations. 

\section*{Acknowledgements}
BS and AH are supported by STFC research grant ST/R000891/1. 
AH is also acknowledges support by STFC Ernest Rutherford Fellowship grant number ST/L00397X/2
We would like to acknowledge and thank the useful discussions with Elena Khomenko, Nikola Vitas, Malcolm Druett and Giulio Del Zanna. The simulation data from this study are available from BS upon reasonable request.


\bibliographystyle{aasjournal} 
\bibliography{losbib} 

\appendix

\section{Ionisation and recombination rates} \label{app:rates}

In dimensional units (of $s^{-1}$), the recombination and ionisation rates are given by the empirical rates in \cite{Voronov1997,Smirnov2003}:

\begin{gather}
    \Gamma _{rec,dim} =  \frac{n_{0} \rho _p}{\sqrt{T_{e0} T_p/T_f}} 2.6\times 10^{-19}, \\
    \Gamma _{ion,dim} =  n_0 \rho _p \frac{2.91 \times 10^{-14}}{0.232+\chi} \chi ^{0.39} e^{\chi}, \\
    \chi = 13.6 \frac{T_f}{T_{e0} T_p}, \\
    T_f = \frac{1}{4} \beta \gamma \frac{2 \xi_{p0}}{\xi_{n0} +2 \xi_{p0}},
\end{gather}
where $n_0$ is a reference dimensional density, and $T_{e0}$ is a reference dimensional temperature in electron volts. The instantaneous plasma temperature $T_p$ is modified by $T_f$ due to the normalisation in this paper being a Alfv\'en speed of unity. At time $t=0$, $T_p/T_f = 1$. 


From the conservation of mass equation, the initial (dimensionless) neutral fraction can be calculated from the ratio of the dimensional rates as:
\begin{gather}
    \xi _n = \Gamma _{rec,dim}/\Gamma _{ion,dim} / (\Gamma _{rec,dim}/\Gamma_{ion,dim} +1), 
\end{gather}
which is a function of normalisation temperature only. Therefore, for a given reference (dimensional) temperature $T_0$, the initial neutral $\xi_{n0}$ and plasma $\xi_{p0}$ fractions can be uniquely determined. 

We normalise the ionisation and recombination rates by the the recombination rate using the normalisation parameters (also in units of $s^{-1}$):

\begin{gather}
    \Gamma _{rec,0} =  \xi_{p0} \frac{n_0}{\sqrt{T_{e0}}} 2.6\times 10^{-19}. \\
\end{gather}

The normalised form of the rates is then given by:
\begin{gather}
    \Gamma _{rec} =  \frac{\Gamma _{rec,dim}}{\Gamma _{rec,0}} \tau = \frac{\rho_p}{\sqrt{T_p}} \frac{\sqrt{T_f}}{\xi _{p0}} \tau _{IR}, \\
    \Gamma _{ion} =  \frac{\Gamma _{ion,dim}}{\Gamma _{rec,0}} \tau = \rho_p \frac{\mbox{e} ^{-\chi} \chi ^{0.39} }{0.232 + \chi} \frac{\hat{R}}{\xi _{p0}} \tau _{IR}, \\
    \hat{R} = \frac{2.91 \times 10 ^{-14}}{2.6 \times 10^{-19}} \sqrt{T_{e0}},
\end{gather}
where $\tau = \tau _{IR}/(\alpha(0) \rho _t)$ is the collisional time multiplied by a factor $\tau _{IR}$ that governs the rate of ionisation and recombination relative to the collisional time. These nondimensional ionisation and recombination rates are independent of the reference density $n_0$ and therefore the reference temperature $T_0$ alone determines the initial rates and equilibrium.

\section{Shock jump derivation} \label{app:iripjump}

The IRIP equations studied in this paper involve a non-conservative energy equation, where the ionisation potential and heating terms provide energy losses and gains from the system. As such, analysing the jump conditions across the shock requires a slightly different set of equations since the upstream energy is not necessarily equal to the downstream energy. Note that momentum and mass across the shock are conserved so these equations have the same jump equations as the MHD case. For the IRIP case we replace the energy equation with Equation \ref{eqn:gmconst} for the balance of the ionisation potential and heating terms in a slightly different form such that the bulk density $\rho=\rho_p+\rho_n$ is used. Choosing a point sufficiently upstream and downstream of the system such that the system is in thermal equilibrium ($T=T_n=T_p)$, ionisation equilibrium ($\Gamma _{rec} \rho_p = \Gamma _{ion} \rho_n$), with zero drift velocity ($\textbf{v} = \textbf{v}_n= \textbf{v}_p$), and a ionisation potential balance ($\phi _I = A_{heat}$), the jump equations in the deHoffman-Teller frame for the IRIP case are as follows:

\begin{gather}
    \left[\rho v_x  \right]^u _d = 0, \label{eqn:appmass} \\
    \left[\rho v_x^2 +P +\frac{B_y^2}{2} \right]^u _d = 0, \\
    \left[\rho v_x v_y -B_x B_y \right]^u _d = 0, \label{eqn:appxym}\\
    \left[B_x \right]^u _d = 0, \label{eqn:appbx}\\
    \left[v_x B_y -v_y B_x   \right]^u _d = 0, \\
    \left[ F(T) \rho ^2 \xi _i ^2  \right]^u _d = 0,
\end{gather}
for upstream (superscript $u$) and downstream (superscript $d$) states. 

It can easily be seen from Equation \ref{eqn:appbx} that $B_x$ is constant across the shock, i.e.,
\begin{gather}
    \frac{B_x^d}{B_x^u}=1.
\end{gather}

By defining a compressional ratio $r$ as
\begin{gather}
    \frac{\rho ^d}{\rho ^u} = r,
\end{gather}
the mass conservation (Equation \ref{eqn:appmass}) gives
\begin{gather}
    \frac{v_x ^d}{v_x ^u} = \frac{1}{r}.
\end{gather}

The $xy$-momentum (Equation \ref{eqn:appxym}) can be written as
\begin{gather}
    \rho^d v_x ^d v_y^d - B_x B_y^d=\rho^u v_x ^u v_y^u - B_x B_y^u, \\
    \frac{\rho^d}{\rho^u} v_x ^d v_y^d - \frac{B_x B_y^d}{\rho^u}=v_x ^u v_y^u - \frac{B_x B_y^u}{\rho^u}, \\
    r v_x ^d v_y^d -\frac{B_x}{\rho^u} B_y^d=\rho^u v_x ^u v_y^u - \frac{B_x}{\rho^u} B_y^u,\\
    v_x ^u v_y^d - \frac{B_x}{\rho^u} B_y^d=\rho^u v_x ^u v_y^u - \frac{B_x}{\rho^u} B_y^u.
\end{gather}
Now, let $B_y=\tan (\theta)$ and define the Alfv\'en Mach number as $V_{Ax}^2= \frac{B_x^2}{\rho ^u}$ to give
\begin{gather}
    v_x^u v_y^d - V_{Ax}^2 \tan (\theta ^d) = v_x^u v_y^u - V_{Ax}^2 \tan (\theta ^u), \\
    v_x^u (v_y^d -v_y^u) = V_{Ax}^2 (\tan (\theta ^d) - \tan (\theta ^u)) .
\end{gather}
In the Hoffman-Teller frame, the electric field is zero, hence $v_y = v_x tan (\theta ^u)$. Also $v_x/v_y=B_x/B_y=1/\tan(\theta)$.
\begin{gather}
    v_x^u (v_x^d \tan (\theta ^d) -v_x^u \tan (\theta ^u)) = V_{Ax}^2 (\tan (\theta ^d) - \tan (\theta ^u)), \\
    v_x^u (v_x^u \tan (\theta ^d)/r -v_x^u \tan (\theta ^u)) = V_{Ax}^2 (\tan (\theta ^d) - \tan (\theta ^u)), \\
    v_x^{2u} (\tan (\theta ^d)/r - \tan (\theta ^u)) = V_{Ax}^2 (\tan (\theta ^d) - \tan (\theta ^u)), \\
    \tan (\theta ^d) \left( \frac{1}{r} - \frac{V_{Ax}^{2}}{v_x^{2u}} \right)=\tan (\theta ^u) \left( 1 - \frac{V_{Ax}^{2}}{v_x^{2u}} \right), \\
    \frac{\tan (\theta ^d)}{\tan (\theta ^u)} = \frac{B_y ^d}{B_y ^u}= r \frac{A_x^{u2} -1}{A_x^{u2} -r}.
\end{gather}

From electric field requirements:
In the deHoffmann-Teller frame the electric field is zero either side of the shock:
\begin{gather}
    B_x=\frac{v_x^d B_y^d}{v_y^d}=\frac{v_x^u B_y^u}{v_y^u}, \\
    \frac{v_y^d}{v_y^u} = \frac{v_x ^d}{v_x ^u} \frac{B_y^d}{B_y^u}, \\
    \frac{v_y^d}{v_y^u} = \frac{A_x^{u2} -1}{A_x^{u2} -r}.
\end{gather}

From the $xx$-momentum equation:
\begin{gather}
    \rho^d v_x^{d2} +P^d +\frac{B_y^{d2}}{2}=\rho^d v_x^{u2} +P^u +\frac{B_y^{u2}}{2}, \\
    \frac{P^d}{P^u}=1+\frac{1}{P^u} \left( \rho^u v_x^{u2} - \rho^d v_x ^{d2} + \frac{B_y^{u2}}{2} - \frac{B_y^{d2}}{2} \right) \\
    = 1 + \frac{1}{P^u} \left( \rho^u \left(v_x ^{u2} - \frac{1}{r} v_x^{u2}\right) \right. \nonumber \\ \left. \hspace{1.0cm} + \frac{B_y^{2u}}{2} \left(1 - r^2 \left(\frac{A_x^{2u}-1}{A_x^{2u}-r}\right)^2\right)  \right).
\end{gather}

The plasma-$\beta$ can be defined as
\begin{gather}
    \beta = \frac{P}{B^2/2}=\frac{2P}{B_x^2+B_y^2}=\frac{2P}{B_x^2(1+\tan ^2 (\theta))}, \\
    P= \frac{B_x^2 \beta (1+\tan(\theta))}{2}.
\end{gather}

Therefore,
\begin{gather}
    \frac{P^d}{P^u} = 1 + \frac{2}{\beta (1+ \tan ^2 (\theta))} \frac{A_x^{u2}}{r} \left[ r-1 +\frac{r \tan ^2 (\theta)}{2 A_x^{u2}} \times \right. \\ \hspace{4.0cm} \left.  \left(1 - r^2 \left(\frac{A_x^{2u}-1}{A_x^{2u}-r}\right)^2\right)  \right].
\end{gather}

The ideal gas law can then be used to relate the upstream and downstream temperatures:
\begin{gather}
    \frac{T^d}{T^u} = \frac{1}{r} \left[1 + \frac{2}{\beta (1+ \tan ^2 (\theta))} \frac{A_x^{u2}}{r} \times \right. \nonumber \\ \hspace{1.0cm} \left. \left( r-1 + \frac{r \tan ^2 (\theta)}{2 A_x^{u2}} \left(1 - r^2 \left(\frac{A_x^{2u}-1}{A_x^{2u}-r}\right)^2\right)  \right) \right]   .
\end{gather}

The upstream and downstream Alfv\'en Mach numbers can be related to the compressional ratio
\begin{gather}
    \frac{A_x^{d2}}{A_x^{u2}}= \frac{\rho^d}{\rho^u}\frac{v_x^{d2}}{v_x^{u2}}=\frac{1}{r}.
\end{gather}

Therefore,
\begin{gather}
    \frac{T^d}{T^u} = \frac{A_x^{d2}}{A_x^{u2}} \left[1 + \frac{2}{\beta (1+ \tan ^2 (\theta))} \left( A_x^{u2}-A_x^{d2} + \frac{\tan ^2 (\theta)}{2} \times \right. \right. \nonumber \\ \left. \left. \hspace{5.0cm} \left(1 - \left(\frac{A_x^{u2}-1}{A_x^{d2}-1}\right)^2\right)  \right) \right]. \label{eqn:apptjumpsf}
\end{gather}

Finally, one can rewrite the ionisation fraction using the recombination and ionisation rates as
\begin{gather}
    \xi _i = \frac{1}{\Gamma _{rec}/\Gamma _{ion} +1} = \chi (T).
\end{gather}
Using this, is is then possible to get an expression relating the temperature jump across a shock to the compressional ratio or the upstream and downstream Alfv\'en Mach numbers:
\begin{gather}
    \frac{F(T ^d)}{F(T^u)} \left( \frac{F(T^u)/G(T^u) +1}{F(T^d)/G(T^d) +1} \right)^2= \frac{1}{r^2} = \frac{A_x^{d4}}{A_x^{u4}}. \label{eqn:appft}
\end{gather}

\begin{figure}
    \centering
\includegraphics[width=0.95\linewidth,clip=true,trim=0.9cm 8.0cm 1.5cm 8.0cm]{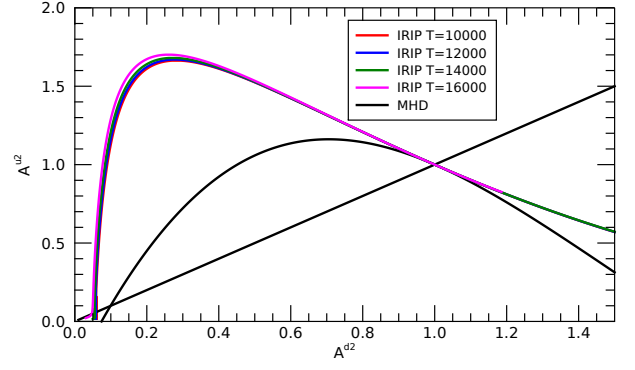}
    \caption{Numerical solutions to the shock jump equations for the MHD (black) and IRIP equations. The trivial solution ($A^{d2}=A^{u2}$) exists for both sets of equations.}
    \label{fig:iripjumps}
\end{figure}

Equations (\ref{eqn:apptjumpsf}) and (\ref{eqn:appft}) can be solved numerically to find possible stable shock solutions to the IRIP equations. Using the upstream properties of the shock in Section \ref{sec:results} ($\beta = 0.13, \theta = 1.16$) the possible shock transitions are shown in Figure \ref{fig:iripjumps}, along with the MHD jump solutions (see \cite{Hau1989}). The MHD solutions hold for the IR case with a conservative energy equation. 

For a switch-off slow-mode shock, the upstream Alfv\'en Mach number is unity. The corresponding downstream Alfv\'en Mach numbers compare well with the simulations, see Figure \ref{fig:mach}. The IRIP solution for the switch-off shock is only stable for a much lower downstream Alfv\'en Mach number than the MHD case. Consequently, the propagation speed of the shock in the IRIP model should be slower than the MHD solution. The IRIP case should also have a much higher compressional ratio across the shock.

\end{document}